\documentclass[fleqn,usenatbib]{mnras}

\usepackage{newtxtext,newtxmath}

\usepackage[T1]{fontenc}
\usepackage{ae,aecompl}

\usepackage{graphicx}	
\usepackage{amsmath}	
\usepackage{cleveref}
\usepackage[nolist,nohyperlinks]{acronym}
\usepackage{booktabs}

\begin{acronym}
    \acro{2fgl}[2FGL]{\textit{Fermi} Large Area Telescope Second Source Catalog}
    \acro{3fgl}[3FGL]{\textit{Fermi} Large Area Telescope Third Source Catalog}
    \acro{4fgl}[4FGL]{\textit{Fermi} Large Area Telescope Fourth Source Catalog}
	\acro{ad}[AD~test]{k-sample Anderson-Darling test}
	\acro{ecdf}[ECDF]{Empirical Cumulative Distribution Function}
	\acro{evpa}[EVPA]{Electric Vector Position Angle}
    \acro{lat}[LAT]{Large Area Telescope}
    \acro{mcoa}[MCoA]{Minimal Change of Angle}
    \acro{mcor}[MCoR]{Minimal Change of Rate}
    \acro{pdf}[PDF]{Probability Density Function}
    \acro{pic}[PIC]{Particle-In-Cell}
    \acro{snr}[SNR]{signal-to-noise-ratio}
	\acro{sps}[SPS]{smoothness +~pairwise significance}
\end{acronym}

\crefname{equation}{Eq.}{Eqs.}
\Crefname{equation}{Equation}{Equations}
\crefname{figure}{Fig.}{Figs.}
\Crefname{figure}{Figure}{Figures}
\crefname{table}{Table}{Tables}
\Crefname{table}{Table}{Tables}
\crefname{section}{Section}{Sections}
\Crefname{section}{Section}{Sections}
\crefname{appendix}{Appendix}{Appendices}
\Crefname{appendix}{Appendix}{Appendices}

\title{The Time-Dependent Distribution of Optical Polarization Angle Changes in Blazars}

\author[S. Kiehlmann et al.]{
S. Kiehlmann,$^{1,2}$\thanks{E-mail: skiehlmann@mail.de}
D. Blinov,$^{1,2,3}$
I. Liodakis,$^{4}$
V. Pavlidou,$^{1,2}$
A. C. S. Readhead,$^{5}$\newauthor
E. Angelakis,$^{6}$
C. Casadio,$^{1,2}$
T. Hovatta,$^{4,7}$
N. Kylafis,$^{1,2}$
A. Mahabal,$^{5}$
N. Mandarakas,$^{1,2}$\newauthor
I. Myserlis,$^{9,10}$
G. V. Panopoulou,$^{5}$
T. J. Pearson,$^{5}$
A. Ramaprakash,$^{8,5,1}$
P. Reig,$^{1,2}$\newauthor
R. Skalidis,$^{1,2}$
A. S\l owikowska,$^{11}$
K. Tassis,$^{1,2}$
J. A. Zensus$^{10}$
\\
$^{1}$Institute of Astrophysics, Foundation for Research and Technology-Hellas, GR-71110 Heraklion, Greece\\
$^{2}$Department of Physics, University of Crete, GR-70013 Heraklion, Greece\\
$^{3}$St. Petersburg State University,Universitetsky pr. 28, Petrodvoretz, 198504 St. Petersburg, Russia\\
$^{4}$Finnish Centre for Astronomy with ESO (FINCA), University of Turku, FI-20014, Turku, Finland\\
$^{5}$Cahill Center for Astronomy and Astrophysics, California Institute of Technology, 1200 E California Blvd, MC 249-17, Pasadena,\\CA 91125, USA\\
$^{6}$Section of Astrophysics, Astronomy \& Mechanics, Department of Physics, National and Kapodistrian University of Athens,\\Panepistimiopolis Zografos 15784, Greece\\
$^{7}$Aalto University, Mets\"ahovi Radio Observatory, Mets\"ahovintie 114, 02540 Kylm\"al\"a, Finland\\
$^{8}$Inter-University Centre for Astronomy and Astrophysics, Post Bag 4, Ganeshkhind, Pune 411 007, India\\
$^{9}$Instituto de Radioastronom\'{i}a Milim\'{e}trica, Avenida Divina Pastora 7, Local 20, E-18012 Granada, Spain\\
$^{10}$Max-Planck-Institut f\"ur Radioastronomie, Auf dem H\"ugel 69, D-53121 Bonn, Germany\\
$^{11}$Institute of Astronomy, Faculty of Physics, Astronomy and Informatics, Nicolaus Copernicus University in Toru\'n, Grudziadzka 5,\\PL-87-100 Toru\'n, Poland \\
}

\date{Accepted XXX. Received YYY; in original form ZZZ}

\pubyear{2021}

\begin{document}
\label{firstpage}
\pagerange{\pageref{firstpage}--\pageref{lastpage}}
\maketitle

\begin{abstract}
At optical wavelengths, blazar \ac{evpa} rotations linked with gamma-ray activity have been the subject of intense interest and systematic investigation for over a decade.
One difficulty in the interpretation of \ac{evpa} rotations is the inherent $180^\circ$~ambiguity in the measurements.
It is therefore essential, when studying \ac{evpa} rotations, to ensure that the typical time-interval between successive observations -- i.e. the cadence -- is short enough to ensure that the correct modulo $180^\circ$ value is selected. 
This optimal cadence depends on the maximum intrinsic \ac{evpa} rotation speed in blazars, which is currently not known.
In this paper we address the following questions for the RoboPol sample: What range of rotation speeds for rotations greater than $90^\circ$ can we expect? What observation cadence is required to detect such rotations? Have rapid rotations been missed in \ac{evpa} rotation studies thus far? What fraction of data is affected by the ambiguity? And how likely are detected rotations affected by the ambiguity?
We answer these questions with three seasons of optical polarimetric observations of a statistical sample of blazars sampled weekly with the RoboPol instrument and an additional season with daily observations.
We model the distribution of \ac{evpa} changes on time scales from 1--30~days and estimate the fraction of changes exceeding $90^\circ$.
We show that at least daily observations are necessary to measure $>96\%$ of optical \ac{evpa} variability in the RoboPol sample of blazars correctly and that intra-day observations are needed to measure the fastest rotations that have been seen thus far.
\end{abstract}



\section{Introduction}
\label{sec:intro}

\citet{2008Natur.452..966M,2010ApJ...710L.126M} reported the first incidents of contemporaneous optical \acf{evpa} rotations and gamma-ray flares.
\citet{blinov_2015,blinov_2018} showed that such contemporaneous events detected in a statistical sample of sources cannot all be explained by chance coincidences; at least some if not all \ac{evpa} rotations have to be physically related to gamma-ray activity and time lags between the two types of events consistent with zero imply co-spatial emission regions. 
Such \ac{evpa} changes of optical polarization could provide a better understanding of the gamma-ray flaring activity in blazars, through (a) revealing a potential physical connection between the optical synchrotron radiation and the high-energy radiation process and (b) elucidating the magnetic field structure of the emitting region.
Various models have been proposed to explain \ac{evpa} rotations.
These include models attributing \ac{evpa} rotations to turbulence \citep[e.g.][]{1985ApJ...290..627J}, or to geometric effects \citep[e.g.][]{2010IJMPD..19..701N,2017MNRAS.467.3876L,2018ApJ...864..140P}.
\citet{2020A&A...636A..79C} introduced a simple, yet versatile two-component model.
Other models have explored the multi-frequency \ac{evpa} changes with a particular focus on optical \ac{evpa} rotations and gamma-ray flares \citep[e.g.][]{2014ApJ...780...87M,2014ApJ...789...66Z}.
Recently \ac{pic} models based on first-principle physics have been introduced \citep{2018ApJ...862L..25Z,2020ApJ...900L..23H}.
These models can be used to constrain assumptions about the magnetic field structure, the jet dynamics, and the radiative processes.

Systematic tests of these models require a representative set of reliably measured rotation events, which is not easily obtained. For example, 
one of the first optical \ac{evpa} rotations reported to coincide with a gamma-ray flare \citep{2010Natur.463..919A} was poorly sampled and later shown to be inconsistent with the originally reported $208^\circ$~rotation \citep[Fig.~2 and~3 by][]{kiehlmann_2016}.

In studying \ac{evpa} rotations one has to be careful that the position angle has not rotated so much between successive observations as to make the $180^\circ$~ambiguity a problem.  The typical time interval between successive observations -- or the cadence\footnote{We use the term \emph{cadence} to refer to the median time interval between successive observations of a source. Thus a ten~day cadence refers to one observation every ten days. A \emph{faster} cadence indicates a shorter time interval between successive observations and a \emph{slower} cadence indicates a longer time interval between successive observations.} -- is therefore critical.  Clearly, if the change in \ac{evpa} between successive observations is $\ll$ $90^\circ$ then this will not be a problem.
One goal of this paper is to estimate the probability of \ac{evpa} changes to exceed $90^\circ$ as a function of cadence.

The  RoboPol project \citep{pavlidou_2014} monitored a sample of 64~gamma-ray loud blazars and a control sample of 13~gamma-ray quiet blazars with an average cadence of 7~days  over three seasons in 2013-2015.
Results from this program were presented and analysed by \citet{angelakis_2016,blinov_2015,blinov_2016a,blinov_2016b, blinov_2018,kiehlmann_2017}.
In 2016 a fourth season of RoboPol observations focused on a smaller sub-sample of sources monitored with faster cadence.
These data enable us to test the effects of the cadence on the analysis of \ac{evpa} rotations and to determine the cadence that is required for such studies.
The distribution of rotation rates -- i.e. the position angle variation per time interval -- enables us to determine the cadence required for accurate determinations of \ac{evpa} rotations. 
\citet{blinov_2016a} discussed the distribution of rotation rates based on the first two seasons of RoboPol data.
With the addition of the fast-cadence data of season~4, we are able to extend the distribution to include significantly faster rotation rates.

This paper is organized as follows. In \Cref{sec:data} we describe the data used in the analysis.
In \cref{sec:ambiguity} we model the \ac{evpa} changes and estimate the fraction of data that is affected by the $180^\circ$~ambiguity at different cadences.
In \cref{sec:randomwalk} we test whether the \ac{evpa} follows a random walk process.
In \cref{sec:samplingdependence} we compare \ac{evpa} rotations identified in seasons~1--3 with season~4 rotations, and test the effects of cadence on the results. In 
\Cref{sec:discussion} we discuss the implications of the effects of cadence and the $180^\circ$~ambiguity on the analysis and interpretation of \ac{evpa} rotations.

\section{Data}
\label{sec:data}

\begin{table}
    \caption{
        Selection criteria for the full sample.
        }
    \label{tab:sample_selection}
    \centering
    \begin{tabular}{p{0.35\columnwidth} c c}
        \toprule
        Property & Main sample & Control sample \\
        \midrule
        \acs{4fgl} & included & not included \\
        \acs{4fgl} $F(E > 100\,\mathrm{MeV})$ & $>10^{-8}\,\mathrm{cm^{-2} s^{-1}}$ & - \\
        \acs{2fgl} source class & agu, bzb, or bzq  & - \\
        Galactic latitude $|b|$ & $>10^\circ$ & - \\
        Altitude (alt) constraints & 
        \begin{minipage}{0.2\columnwidth}\begin{center}$\mathrm{alt}_\mathrm{max} \geq 40^\circ$  \\ Jun--Nov$^*$\end{center}\end{minipage} & 
        \begin{minipage}{0.2\columnwidth}\begin{center} $\mathrm{alt}_\mathrm{max} \geq 40^\circ$  \\ Apr--Nov$^*$\end{center}\end{minipage} \\
        R magnitude & $\leq17.5$ & $\leq17.5$ \\
        \raggedright CGRaBS/15~GHz OVRO monitoring  & No constraints & Included \\
        \raggedright OVRO 15~GHz mean flux density & No constraints & $\geq0.060\,\mathrm{Jy}$ \\
        \raggedright OVRO 15~GHz intrinsic modulation index, $m$ & No constraints & $\geq0.05$ \\
        \bottomrule
        \multicolumn{3}{p{.9\columnwidth}}{
            $^*$ Constraint on the sources' maximum altitude at Skinakas observatory for at least 90~consecutive nights in the stated time window.
        } \\
    \end{tabular}
\end{table}{}

 The data analysed in this work were obtained with the RoboPol instrument \citep{2019MNRAS.485.2355R} at the 1.3~meter telescope of the Skinakas observatory in Crete, Greece.
The complete set of RoboPol blazar data is described and published by \citet[in the following referred to as data release (DR) paper]{2020MNRAS.tmp.3639B}.
We selected the same 77~sources from the samples of main and control sources presented in the DR~paper that were analysed by \citet{blinov_2018}.
This sample has been selected on the basis of stringent, objective, and bias-free criteria.
The selection criteria are listed in \cref{tab:sample_selection}.
These criteria and the corresponding statistically complete sample were initially described by \citet{pavlidou_2014}.
From the parent sample 62~main sample sources and 15~control sample sources were randomly drawn.
We note that the \textit{Fermi}-\ac{lat} catalog associations were initially taken from the \ac{2fgl}.
As explained by \citet{blinov_2016b} two of the initial control sample sources were moved to the main sample after the release of the \ac{3fgl}.
In the following we refer to these 77~sources as the \emph{full sample}, which was observed during seasons~1--3 (2013--2015).

From the season~1--3 data we calculated the \ac{evpa} rate of change, i.e. the absolute change of \ac{evpa} divided by the time that elapsed between observations for each pair of successive data points.
For each source we calculated the median of the \ac{evpa} rates of change and selected the 29~main and control sample sources with the largest median rates.
Of those sources, RBPLJ1653+3945 was excluded because of calibration problems.
In the fourth RoboPol season (2016) the resulting sub-sample of 28~sources was observed at a faster cadence to determine how a faster cadence would impact the results.

In the following we refer to the above 28 sources as the \emph{season~4 sample}.
While the full sample is bias-free within the constraints given by the selection criteria (\cref{tab:sample_selection}), the season~4 sample is biased towards rapid changes of the \ac{evpa} due to the selection criteria.
\cref{tab:sources} lists the \emph{full sample} of RoboPol sources considered in the analysis and indicates the season~4 sub-sample (`S4' in the last column).

When we characterize the variability of the \ac{evpa} and its dependence on the time separation in the following analysis, we do not distinguish between main and control sample sources but combine them jointly in the full and the season~4 sample.
We chose to do this to increase the number of data points for the statistical analysis.
However, we note that the results will relate only to this specific selection and combination of sources and may not be generally extended to other samples of blazars.

\begin{table}
    \caption{RoboPol sources used in this publication: RoboPol source name (col. 1), J2000 right ascension and declination (col. 2+3), `S4' marks sources that were observed during seasons~1--4 (col. 4) other sources were observed during seasons~1--3 only. The full table is available online.}
    \label{tab:sources}
    \begin{tabular}{l r r r}
        \toprule
        RoboPol source name & RA [h:m:s] & Dec [d:m:s] & Season \\
        \midrule
        RBPLJ0017+8135 &       00:17:08 &       +81:35:08 &     \\
        RBPLJ0045+2127 &       00:45:19 &       +21:27:40 & S4  \\
        RBPLJ0114+1325 &       01:14:53 &       +13:25:37 & S4  \\
        RBPLJ0136+4751 &       01:36:59 &       +47:51:29 & S4  \\
        RBPLJ0211+1051 &       02:11:13 &       +10:51:35 & S4  \\
        \dots \\
        \bottomrule
    \end{tabular}
\end{table}

\begin{figure}
  \centering
  \includegraphics[width=\columnwidth]{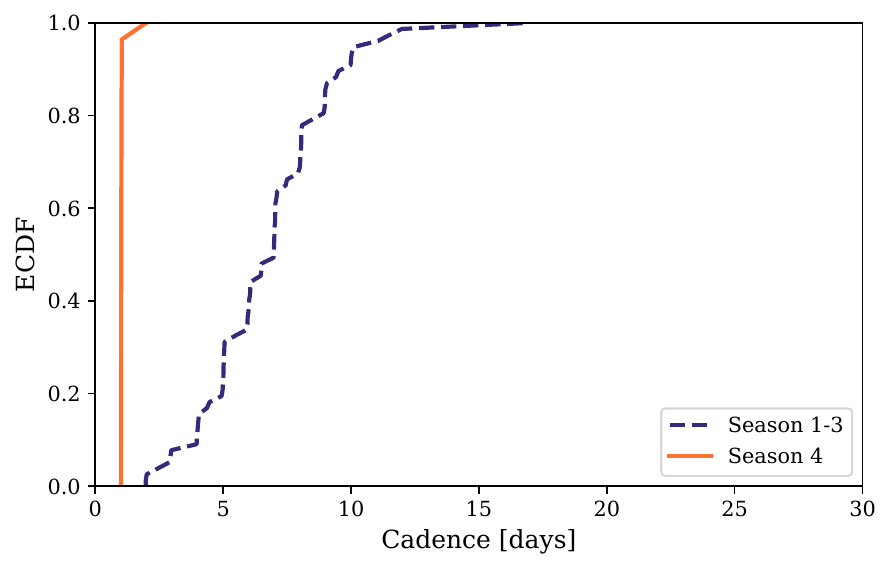}
  \caption{
    \acf{ecdf} of the cadence of observations over all sources in different seasons: The full sample seasons~1--3 data (purple, dashed) compared with the season~4 data (orange, solid). The cadence in season~4 was 1~day, the median cadence in seasons~1--3 was 7~days.}
   \label{fig:timesampling}
\end{figure}

\Cref{fig:timesampling} shows the \acp{ecdf} of the cadence at which sources were observed during seasons~1--3 and season 4.
On average the cadence is about 7~times faster for season~4.
In the following we test how the faster cadence affects the identification of \ac{evpa} rotations in season~4.

\section{Cadence and the 180 degree ambiguity}
\label{sec:ambiguity}

The \ac{evpa}, $\chi$, is measured in an interval of $180^\circ$.
The total amount of change\footnote{We use the term \emph{\ac{evpa} change}, when we refer to a difference of the \ac{evpa} between two measurements. We do not use the term \emph{rotation} to avoid confusion with its common use for \emph{rotation events}, where the \ac{evpa} gradually and smoothly changes in the same direction for a period of time sampled with multiple data points.} between two measurements is not uniquely established, because the change may have been the measured difference, $\Delta\chi$, plus an unknowable integer multiple of $180^\circ$.
This is the so-called \emph{$180^\circ$~ambiguity} or \emph{$n\pi$ ambiguity}.
In this section we estimate the extent to which the measured data are affected by the $180^\circ$~ambiguity.
We start with the introduction of three terms, the \emph{intrinsic}, the \emph{adjusted}, and the \emph{wrapped \ac{evpa} change}.

\begin{figure}
  \centering
  \includegraphics[width=\columnwidth]{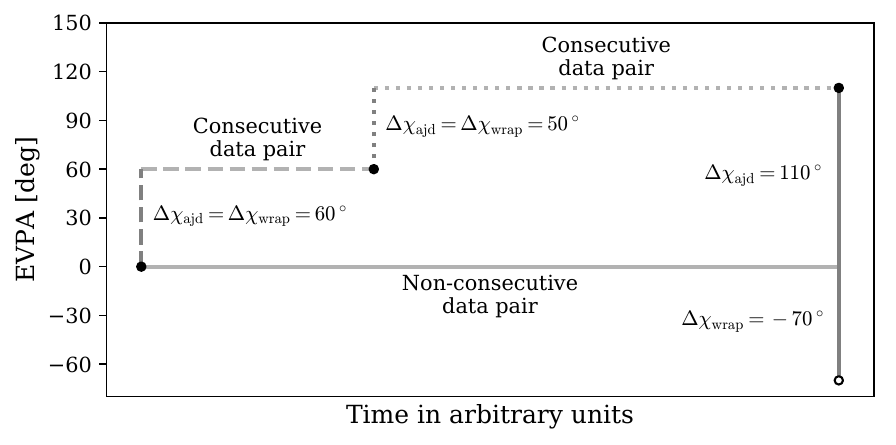}
  \caption{
    Sketch illustrating adjusted (filled circles) and wrapped (open circle) \ac{evpa} changes between consecutive and non-consecutive data pairs.}
  \label{fig:sketch}
\end{figure}

At any two moments in time, $t_1,t_2$, we can measure the \ac{evpa}, $\chi_1,\chi_2$.
The measured change of the \ac{evpa}, $\Delta\chi_\mathrm{meas} = \chi_2 - \chi_1$ is ambiguous, because every change of $\Delta\chi_\mathrm{intr} = \Delta\chi_\mathrm{meas} \pm n \times 180^\circ$ results in the same measurement, with $n \in \mathbb{N}_0$, the set of positive integers.
Here, $\Delta\chi_\mathrm{intr}$ is the \emph{intrinsic change}, i.e. the actual amount by which the \ac{evpa} changed.
Strictly speaking the intrinsic change cannot be determined with certainty from the measurements  without {\it continuous\/} \ac{evpa} observations, due to the $180^\circ$~ambiguity. However provided the change in intrinsic \ac{evpa} between successive observations is $\ll90^\circ$, we \textit{can} determine the change in intrinsic \ac{evpa} between successive, discontinuous observations.

The \emph{adjusted \ac{evpa} change}, $\Delta\chi_\mathrm{adj}$, aims at reproducing the intrinsic \ac{evpa} progression.
This is commonly used in the literature \citep[e.g.][]{kiehlmann_2017,2018ApJ...862....1C,2018A&A...619A..45M}.
We assume that the \ac{evpa} changed minimally between successive measurements; an alternative assumption is discussed in \cref{app:adjustment}.
Under this assumption we pick the \ac{evpa} change with the smallest absolute value in the $(-90^\circ, +90^\circ]$-interval for consecutive data points.
As such, each data point, $\chi_i$, is adjusted relative to its preceding data point, $\chi_j=\chi_{i-1}$, as follows, where $\mathrm{round(\dot)}$ denotes rounding to the nearest integer:

\begin{align}
    \chi_{i,\mathrm{adj}} = \chi_i - n\times 180^\circ \text{\ \ with\ \ } n = \mathrm{round}\left( \frac{\chi_i - \chi_{j}}{180^\circ} \right)
    \label{eq:adjustment}
\end{align}

In \cref{fig:sketch} the black dots illustrate an adjusted \ac{evpa} curve, where the first pair changed by $60^\circ$, the second by $50^\circ$, which results in an adjusted change of $110^\circ$ between the first and third data point.
Adjusted \ac{evpa} changes between consecutive data points are always in the interval $(-90^\circ, +90^\circ]$; for non-consecutive data pairs the adjusted change can exceed this interval in both directions.
Whether an adjusted \ac{evpa} curve correctly represents the intrinsic \ac{evpa} progression, depends on the cadence.
Without any known physical constraints on how fast the \ac{evpa} can rotate in blazars, we cannot know a~priori what cadence is required to reconstruct the intrinsic behaviour correctly from the data.

We introduce the \emph{wrapped \ac{evpa} change}, $\Delta\chi_\mathrm{wrap}$,  as a concise way of expressing \ac{evpa} changes on all time-scales.
For any data pair, $(\chi_i, \chi_j)$ with $j{<}i$ -- whether consecutive or not -- we shift, $\chi_i$ according to \cref{eq:adjustment}, before we calculate the difference between the two values to obtain the wrapped \ac{evpa} change.
The wrapped change between any two measurements is in the interval $(-90^\circ, +90^\circ]$.
For consecutive data points the wrapped change equals the adjusted one.
For non-consecutive data points the wrapped change is the value that would be measured as the adjusted change if no other measurements were taken in-between.
For non-consecutive data points the wrapped and adjusted change may differ, as illustrated in \cref{fig:sketch} for the first and third data point.
The wrapped change is defined between individual data pairs and cannot be used to construct an \ac{evpa} curve with multiple data points.
It is not aimed at reconstructing the intrinsic \ac{evpa} progression.
Instead, we will use the wrapped \ac{evpa} changes to model the distribution of intrinsic \ac{evpa} changes on a statistical basis.

With these definitions of \ac{evpa} changes, we may now describe our statistical treatment of the data.
For each source we consider the \ac{evpa} measurements as a function of time and construct its adjusted \ac{evpa} curve.
For each measurement at time $t_i$ we calculate the adjusted and wrapped \ac{evpa} change with all points at times $t_j$, $j > i$.
The time interval $t_j-t_i$ is registered, and we refer to it as separation.\footnote{We use the term \emph{separation} to distinguish it from the cadence. Note that for a particular source in a particular season the cadence is fixed but the separation ranges from the time between the closest two observations to the most widely separated two observations.}
Like any angle difference, the \ac{evpa} changes are signed, and can take both positive and negative values.
However, the sign is not of relevance to our investigation here.
By construction, the wrapped change does not contain information about the direction of the intrinsic \ac{evpa} change and the adjusted \ac{evpa} changes are as likely to be positive as to be negative.
We therefore only use absolute values for the adjusted and wrapped \ac{evpa} changes.

In the following we propose a model for the distribution of wrapped \ac{evpa} changes that enables us to  model the distribution of intrinsic \ac{evpa} changes.
We will then compare the inferred distribution of intrinsic \ac{evpa} changes to the measured distribution of adjusted \ac{evpa} changes to test how reliable the method of adjusting the \ac{evpa} curve is in reconstructing the intrinsic \ac{evpa} progression for various separations.

\subsection{Model description}
\label{sec:model}

\begin{figure}
  \centering
  \includegraphics[width=\columnwidth]{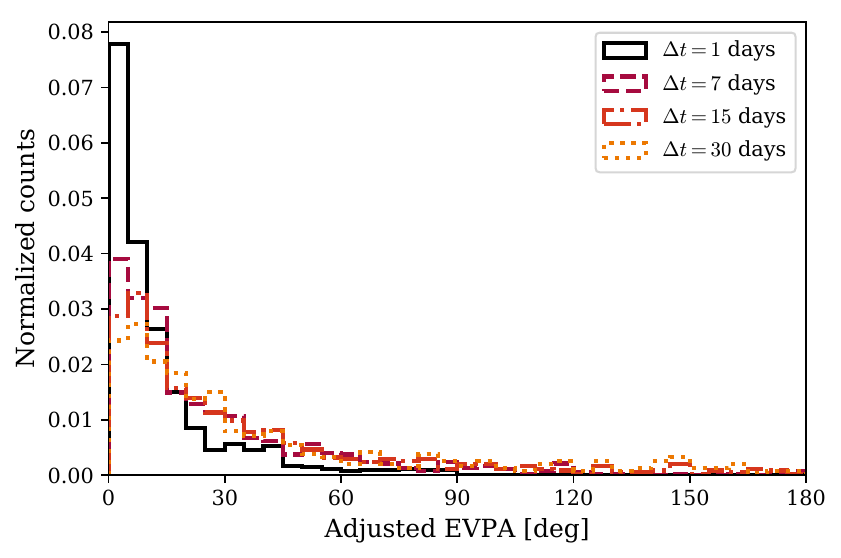}
  \caption{
    Normalized histogram of the adjusted \ac{evpa} changes for different time separations as stated in the legend, $\pm 0.5$~days.}
  \label{fig:wrappedevpadist}
\end{figure}

The \ac{pdf} of adjusted \ac{evpa} changes shows a flattening and a shift in the peak of distributions towards larger \ac{evpa} changes for increasing time separations (\cref{fig:wrappedevpadist}).
These \acp{pdf} resemble the behaviour of a log-normal distribution, where the mean of variable's natural logarithm depends on the time separation.
However, in particular for longer separations we expect that the distribution is biased due to the $180^\circ$ ambiguity and we aim to reconstruct the distribution of intrinsic \ac{evpa} changes in the following.
Motivated by this observation, we assume that the distribution of intrinsic \ac{evpa} changes follows a log-normal distribution, $\mathrm{PDF}_\mathrm{intr}(\Delta\chi_\mathrm{intr};\mu,\sigma) = \mathcal{LN}(\Delta\chi_\mathrm{intr};\mu,\sigma)$, where the natural logarithm of the variable has the mean, $\mu$, and standard deviation, $\sigma$.
We discuss the implications of this assumption at the end of this section.
The absolute intrinsic \ac{evpa} changes can take any values larger than zero.
Intrinsic \ac{evpa} changes exceeding $90^\circ$ are wrapped back into the $[0^\circ, 90^\circ]$-interval when measured as wrapped \ac{evpa} changes.
In \cref{app:model} we show that if the \ac{pdf} of intrinsic changes is log-normal, the wrapped changes can be described by a modified log-normal distribution,  $\mathrm{PDF}_\mathrm{wrap}(\Delta\chi_\mathrm{wrap};\mu,\sigma)$, with parameters $\mu$ and $\sigma$ derived in the appendix.
Through fitting the measured distribution of wrapped \ac{evpa} changes, we can infer the parameters $\mu,\sigma$ of the distribution of intrinsic \ac{evpa} changes.

The best-fit values for parameter, $\mu$, depend on the separation, $\Delta t$ (c.f.~ \cref{app:model}).
We choose $\mu(\Delta t) = \beta_0 + \beta_1 \log(\Delta t)$, where $\beta_0, \beta_1$ are free model parameters.
We find that the standard deviation, $\sigma$, is independent of the separation and include it as free parameter in the model.
We implement the model in \verb|pystan|\footnote{\url{https://pystan.readthedocs.io/}}, a \verb|python| interface to the Bayesian modelling language \verb|Stan|\footnote{\url{https://mc-stan.org/}}.
We use diffuse priors for the model parameters $\beta_0,\beta_1,\sigma$.

For fitting the model to the measured wrapped \ac{evpa} changes we consider all data pairs -- consecutive and non-consecutive -- from the full sample up to a separation of 30~days, giving a total of 19\,585~wrapped \ac{evpa} changes. The inferred parameters with $1\sigma$-credible intervals are: $\beta_0 = 2.43 \pm 0.03~[\ln(\mathrm{deg})]$, $\beta_1 = 0.28 \pm 0.01~[\ln(\mathrm{deg})/\ln(\mathrm{days})]$, $\sigma=1.04 \pm 0.01~[\ln(\mathrm{deg})]$.
Examples of comparisons between the wrapped model and data for different separations are shown in \cref{app:model}.
This model allows us to estimate the expected variability of the \ac{evpa} on different time scales in a sample of blazars.
We note that this model is only informed by data with separations from 1--30~days and the extrapolation towards shorter or longer time scales may not be applicable.
In the following we use the model to estimate the amount by which the adjusted \ac{evpa} curves fail to reproduce the intrinsic \ac{evpa} changes.

We note that the model depends on the assumption that the distribution of the intrinsic \ac{evpa} over the full sample follows a log-normal distribution on all time scales.
This assumption is motivated by the observations discussed above.
We caution the reader that all results that are based on this model depend on this assumption.
Future observations and physical model simulations may allow us to better select and constrain a distribution model.

\subsection{Comparison of intrinsic and adjusted EVPA changes}
\label{sec:modelresults}

\begin{figure*}
  \centering
  \includegraphics[width=\textwidth]{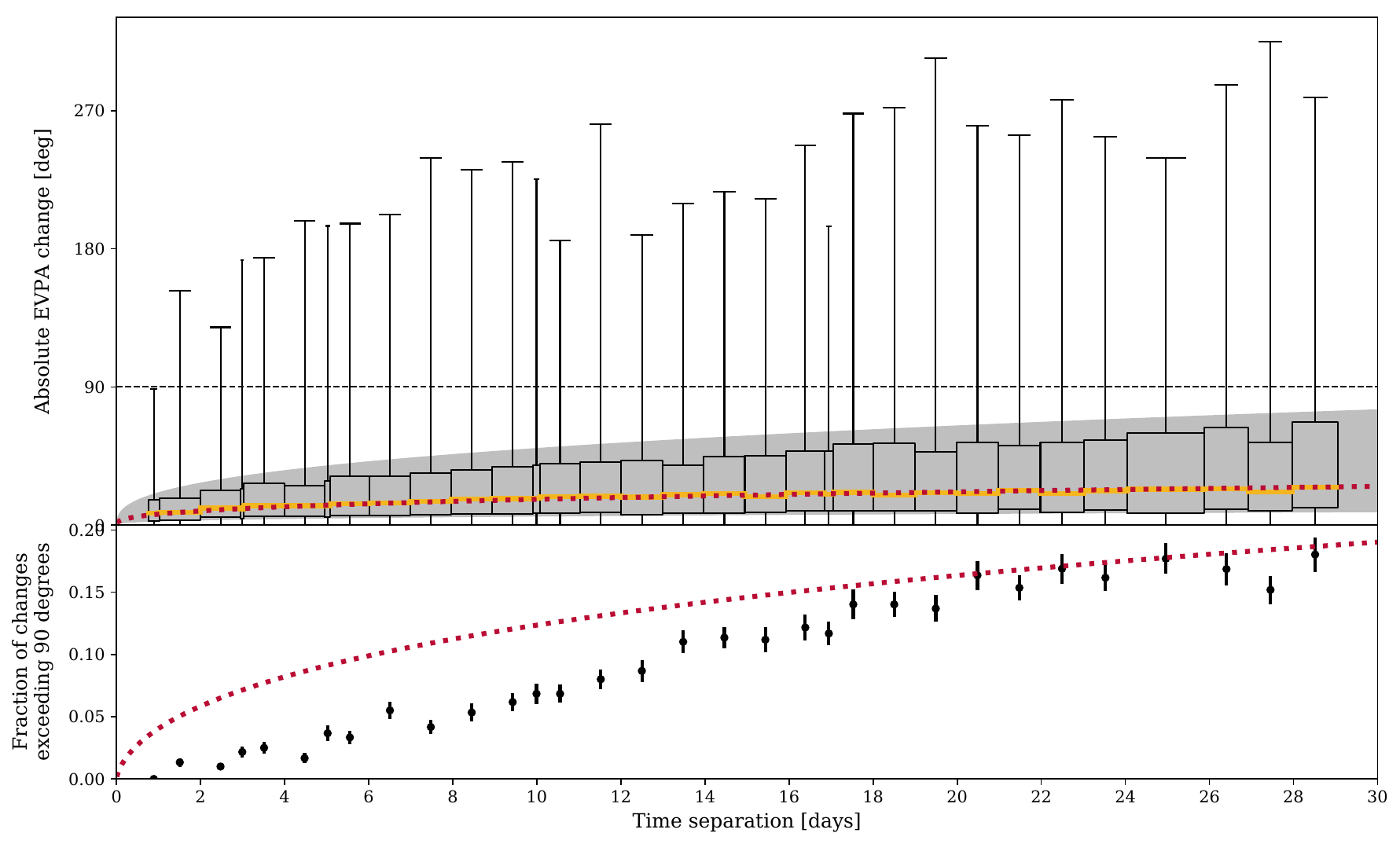}
  \caption{
    Top panel: The boxplots show the distributions of adjusted \ac{evpa} changes, $\Delta\chi_\mathrm{adj}$, at different separations for the full RoboPol sample. Boxes correspond to the 25- and 75-percentile, yellow bars to the median, and lower and upper bars to the minimum and maximum values. The time-scale bins are chosen adaptively such that each box is based on 600~data points. The red, dotted line marks the median of the model distribution, $\Delta\chi_\mathrm{intr}^\mathrm{model}$. The grey region highlights the 25- to 75-percentile region of the model distribution.The horizontal, dashed line highlights $90^\circ$.
    Bottom panel: Fraction of \ac{evpa} changes that exceed $90^\circ$. Black dots represent real measurements from the adjusted \ac{evpa} changes. The corresponding uncertainties are estimated with a bootstrap method; in 100~iterations a random sample of \ac{evpa} changes is selected and the analysis is repeated on this selection; the uncertainty is given by the standard deviation over all bootstrap iterations. The red, dotted line shows the model expectation.}
  \label{fig:evpavartimescale}
\end{figure*}

\Cref{fig:evpavartimescale} shows various percentiles of the distribution of adjusted \ac{evpa} changes for different separations in comparison to the expectation of the intrinsic distribution estimated from the model fit to the wrapped \ac{evpa} changes.
In the upper panel we show the measured distributions of adjusted \ac{evpa} changes for different separations.
As can be seen there, on all separations $>1$~day we find examples of \ac{evpa} changes exceeding $90^\circ$.
The adjusted \ac{evpa} changes generally increase towards larger separations.
Therefore, the fraction of \ac{evpa} changes that exceed $90^\circ$ increases as well, as is shown in the bottom panel.
We note that we are only able to measure \ac{evpa} changes $>90^\circ$ in the adjusted \ac{evpa} when we have more than two data points.

The 25-percentile and the median of the adjusted \ac{evpa} changes and the model of the intrinsic changes are in good agreement over all tested separations.
This shows that the lower half of adjusted \ac{evpa} changes is not strongly affected by the $180^\circ$ ambiguity and that smaller adjusted changes reliably reproduce the intrinsic \ac{evpa} changes.
The 75-percentile of the model distribution suggests that the intrinsic distribution of \ac{evpa} changes has a more extended tail at high values than we find in the adjusted data.
Consequently we find that the measured fraction of adjusted \ac{evpa} changes exceeding $>90^\circ$ is smaller than expected from the intrinsic distribution model (bottom panel).
This discrepancy can be explained by the fact that we cannot measure \ac{evpa} changes larger than $>90^\circ$ between consecutive data points.
\ac{evpa} changes larger than $>90^\circ$ are incorrectly measured and appear as \ac{evpa} changes smaller than $90^\circ$, and this biases the distribution of adjusted \ac{evpa} changes towards smaller values.

Two main conclusions can be drawn from \cref{fig:evpavartimescale}.
First, on all separations longer than 1~day we find \ac{evpa} changes exceeding $>90^\circ$ (upper panel).
Second, on all separations the discrepancy between the expected and the measured fraction of \ac{evpa} changes exceeding $90^\circ$ shows that a fraction of our data is affected by the $180^\circ$~ambiguity and therefore that some of the adjusted \ac{evpa} curves do not correctly represent the intrinsic variability. We observe that the discrepancy (i.e. the difference between the red dashed line and the data in the lower panel of  \cref{fig:evpavartimescale}) first increases with increasing separation and then decreases towards a separation of about 20~days, above which there is no significant discrepancy.
Most of the data (seasons~1--3) were sampled with an average cadence of 7~days, which means that typically only two data points are available on the time scale of 7~days to estimate the changes. With only two data points we are not able to detect any intrinsic \ac{evpa} changes  $>90^\circ$.
The \ac{evpa} changes exceeding $90^\circ$ that we find on this timescale arise either from a (rare) faster cadence in seasons~1--3 or from the season~4 data, when more than two data points are available.
Therefore, at time separations of 7~days the data is mostly sampling-limited.
Because the season~1--3 observations dominate on the separation of about 7~days, here, the discrepancy between expectation and observation is largest.
Towards shorter separations, two effects reduce the discrepancy. First, the \ac{evpa} changes decrease towards shorter separations (\cref{fig:evpavartimescale}, upper panel). Therefore, the fraction of data exceeding $90^\circ$ decreases. Second, these separations are mostly from the season~4 observations, which had an average cadence of one day. Therefore, the \ac{evpa} changes on the shortest separations ($\lesssim7$~days) are better sampled than larger separations.
On longer separations ($\gtrsim 7$~days) the discrepancy is also gradually reduced due to the combination of two effects.
First, the \ac{evpa} changes do not increase linearly with the separation as seen in the upper panel of \cref{fig:evpavartimescale}.
Second, on longer separations we have multiple data points to sample the \ac{evpa} changes, e.g. on 14~days separation typically three data points sample the \ac{evpa} changes, which allows us to detect at least some of the \ac{evpa} changes exceeding $90^\circ$. However, this does not imply that an \ac{evpa} curve is more accurate on longer separations than on shorter separations: this is only the case for the subset of events for which all \ac{evpa} changes sampled by consecutive measurements were smaller than $90^\circ$. In contrast,  if the adjustment of \ac{evpa} data fails on short separations, the adjusted curve will not represent the intrinsic behaviour correctly on longer separations either.
The results  demonstrate that on a statistical basis we sample the distribution of \ac{evpa} changes more accurately on longer separations (>20~days) than on the shorter ones, where the observations are sampling limited.

The model allows us to estimate the fraction of data points that would be affected by the $180^\circ$~ambiguity and thus would incorrectly represent the intrinsic \ac{evpa} changes, for a given cadence.
At the median cadence of seasons~1--3 (7~days) we see from the lower panel of \cref{fig:evpavartimescale} that we expect 11\% of the \ac{evpa} changes to be affected by the $180^\circ$~ambiguity, and that at the median cadence of season~4 (1-day) the fraction  drops to 4\%, i.e. a factor of 2.8~improvement.

The fastest \ac{evpa} change in the joint seasons~1--3 and season~4 data was measured in RBPLJ2253+1608 at JD~2456887.3 (season~2) with a $73^\circ$ change over 18~hours, corresponding to a rate of $96\pm5^\circ/\mathrm{day}$.
To avoid under-sampling of such fast \ac{evpa} changes, we should observe a source at least once every 22~hours.

\section{Random-walk EVPA changes in the intrinsic EVPA?}
\label{sec:randomwalk}

In \cref{sec:model}, we estimated the intrinsic distribution of \ac{evpa} changes for all separations through fitting a model to the observable wrapped \ac{evpa} changes. We can also use the observed distributions of wrapped \ac{evpa} changes on different timescales to test whether the long-term \ac{evpa} changes are a result of independent short-term \ac{evpa} changes, i.e. whether it can be described as a random walk. 
To this end, we construct simulated \ac{evpa} curves based on two assumptions. 
The first assumption is that the \ac{evpa} changes on the shortest separation, $1\pm0.5$~days, is measured correctly (i.e. that the intrinsic $\Delta \chi$ for pairs separated by $\sim$1~day do not exceed $90^\circ$, and that they can therefore be correctly measured from the adjusted \ac{evpa} curves).
Our results from \cref{sec:modelresults} indicate that only 4\% of the data are expected to be incorrectly measured at this separation, and hence our assumption is reasonable.
The second assumption is that the long-term \ac{evpa} changes are a result of independent short-term changes, i.e. they can be described as a random walk in $\Delta\chi$.
We now test this assumption. 

We produce simulated \ac{evpa} random walks as follows:
From the observed distribution of \ac{evpa} changes, $\Delta \chi$, on our shortest cadence (1~day), we randomly draw 200~$\Delta \chi$.\footnote{ 
200~data points with 1~day separation are sufficient to cover the longest observing period in our data.}
The estimated probability of a sign change between two consecutive data pairs over our whole sample is 55\%. 
Therefore, we randomly assign sign \ac{evpa} changes to the drawn $\Delta \chi$ according to a binomial distribution with success probability $p=0.55$. We use these 200~signed $\Delta \chi$ to produce a simulated \ac{evpa} curve. We repeat the process 1000~times, and produce 1000~distinct simulated \ac{evpa} curves. We then measure the wrapped \ac{evpa} changes, $\Delta\chi_\mathrm{wrap}$, on various timescales, from our simulated curves. The wrapped \ac{evpa} changes -- as measured in both the observed data and the simulations -- are unambiguously defined.

\begin{figure}
  \centering
  \includegraphics[width=\columnwidth]{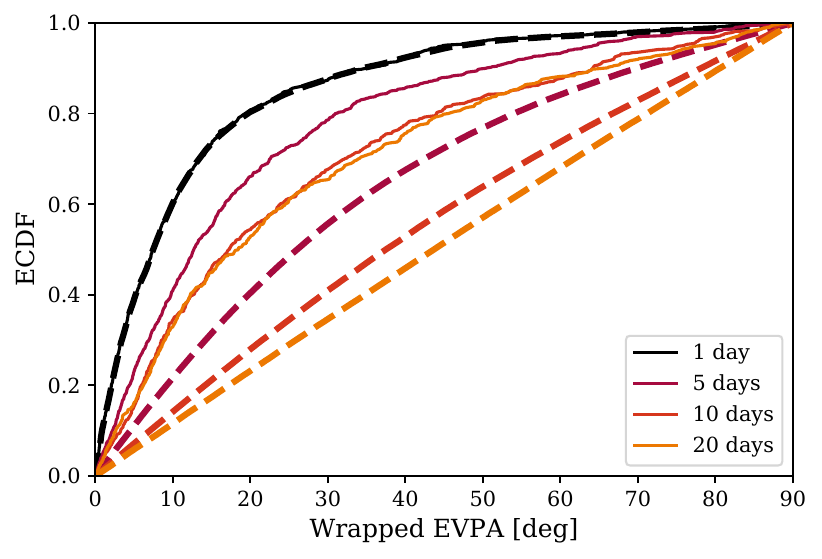}
  \caption{\ac{ecdf} of wrapped \ac{evpa} changes. Different colours represent different cadences, which are shown in the legend. The uncertainties in the cadences are $\pm0.5$~days. Solid lines represent measured data. Dashed lines represent the random walk simulations.
  }
  \label{fig:randomwalk}
\end{figure}

The observed and simulated distributions of $\Delta\chi_\mathrm{wrap}$ on a 1~day cadence will match by construction, since the simulated $\Delta\chi_\mathrm{wrap}$ are directly drawn from the observed distribution. If our second assumption above holds, i.e. the long-term \ac{evpa} changes are a result of independent short-term \ac{evpa} changes, then the  distributions of $\Delta\chi_\mathrm{wrap}$ on longer timescales in the simulations should also match the observed ones. 
\Cref{fig:randomwalk} shows the distributions of $\Delta\chi_\mathrm{wrap}$ from the simulations, together with those we observed. As expected, on a 1~day cadence the distributions match perfectly. However, for longer cadences the simulation-based distributions converge to a uniform distribution, and  differ significantly from the observations. In other words, long-term \ac{evpa} changes introduced by successive, random, short-term \ac{evpa} changes \emph{strongly exceed} the \ac{evpa} changes that we observe for corresponding cadences.
We therefore conclude that the long-term \ac{evpa} changes are not simply a result of random, short-term \ac{evpa} changes.
This is consistent with the finding from our analyses of seasons 1--3 that the \ac{evpa} changes observed over the entire RoboPol sample cannot be attributed solely to \ac{evpa} random walks \citep{blinov_2015, kiehlmann_2017}.
This analysis is based on sample statistics and its results may not be applicable to individual sources.

\section{The identification of rotations}
\label{sec:samplingdependence}

In this section we estimate the effects of cadence and the $180^\circ$ ambiguity on the identification of \ac{evpa} rotations.
To identify \ac{evpa} rotations in our data, we use a method based on \citet{blinov_2015}.
The following requirements must be met in order for a set of measurements to be identified as a smooth \ac{evpa} rotation: 

\begin{enumerate}
    \item The \ac{evpa} has to change consistently in one direction and the rotation rate must not change by more than a threshold value, chosen to be a factor of~5 from the previous measurement, as originally introduced by \citet{blinov_2015}.
    \item The \ac{evpa} has to change by at least $90^\circ$ between first and last measurement.
    \item The \ac{evpa} difference between the first and last data point has to be significant compared to measurement uncertainties.
    \item The rotation has to be sampled by at least four measurements. 
\end{enumerate}

For criterion~(iii) \citet{blinov_2015,blinov_2016a,blinov_2016b,blinov_2018} required that each pair of consecutive data points shows a significant difference.
However, eventually point-to-point \ac{evpa} changes will stop being significant as the \ac{evpa} curve sampling becomes denser at constant measurement uncertainties.
Keeping the consecutive-point-significance requirement would then result in spuriously dismissing rotations.
Therefore, in this work we relaxed this requirement to significance between the first and last data point only.

We consider gaps longer than 30~days between consecutive measurements to automatically break a rotation.
This last criterion only affects season~1--3 data, as season~4 was observed continuously without long gaps.
We call each period of consecutive data points that are separated by less than 30~days an \emph{observing period}.

The difficulties encountered in the determination of \ac{evpa} rotations in blazars are clearly either intrinsic to the process or extrinsic. The only intrinsic difficulty is the $180^\circ$~ambiguity. The extrinsic difficulties are caused by sensitivity limitations of our instruments, cadence, and our choice of parameters in identifying rotations.  We discuss the extrinsic difficulties in \cref{app:extrinsic}, and focus, for the rest of this paper on the $180^\circ$ ambiguity and our scientific findings.

\subsection{EVPA adjustment}
\label{app:adjustment}

Before the analysis, the measured \ac{evpa} curve is typically adjusted under the assumption of a \ac{mcoa} between all pairs of consecutive data points \citep[e.g.][]{kiehlmann_2016}, i.e. data points are shifted by multiples of $\pm 180^\circ$, such that the difference between the shifted data point and its preceding data point is minimal, c.f.~\cref{eq:adjustment}.

\begin{figure}
  \centering
  \includegraphics[width=\columnwidth]{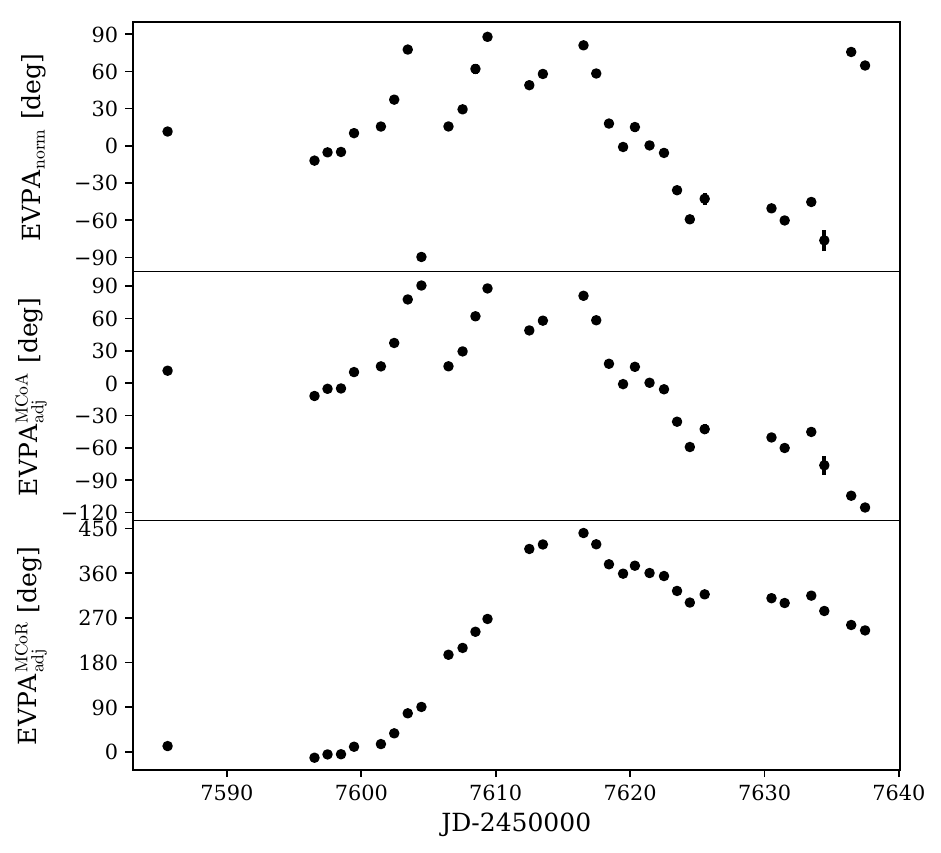}
  \caption{
    Top panel: Season~4 \ac{evpa} data of RBPLJ2202+4216 measured in the [-90, 90] degrees interval. Mid panel: \ac{evpa} data adjusted by the \ac{mcoa} method. Bottom panel: \ac{evpa} data adjusted by the \ac{mcor} method.}
  \label{fig:adjustment}
\end{figure}

The season~4 observations of RBPLJ2202+4216 shown in \cref{fig:adjustment} indicate that the \ac{evpa} progression frequently changed sign between JD~2457595 and JD~2457617.
However, three periods of continuous rotations in the same direction with two larger gaps allow the interpretation that this whole period is one long rotation in the same direction.
Motivated by this example, we explore a second method that assumes a \ac{mcor}.
First, we estimate the rotation rate between two data points, $(t_{i-1}, \chi_{i-1})$, $(t_i, \chi_i)$ through $\dot{\chi}_i = (\chi_i - \chi_{i-1}) / (t_i - t_{i-1})$.
Then we shift data point $i$ by multiples of $\pm 180^\circ$, such that the difference between $\dot{\chi}_i$ and $\dot{\chi}_{i-1}$ becomes minimal.
The second data point in the time series is shifted according to the \ac{mcoa} method.

Both methods fail to reconstruct the intrinsic \ac{evpa} curve when the data are critically under-sampled, but the conditions for this to happen differ.
\ac{mcoa} fails when the intrinsic change between two data points exceeds $90^\circ$.
\ac{mcor} fails when the intrinsic rate is faster than the estimated rate.
We test both methods on random walk simulations based on the model described by \citet{kiehlmann_2016}.
We showed in \cref{sec:randomwalk} that the \ac{evpa} progression of blazars does not follow a random walk.
However, random walks mimic \ac{evpa} changes in blazars well enough to test the two adjustment methods on such simulations.
The model consists of multiple cells with randomized magnetic field orientation. 
At each time step, one cell changes its orientation.
We resample the simulated \ac{evpa} curve to a slower ``observing'' cadence and reduce the ``observed'' angles to the ``measured'' $180^\circ$ range.
Finally, we use the \ac{mcoa} and \ac{mcor} method to adjust the \ac{evpa} curve and cross-check the result with the intrinsic curve.
For various simulation setups (different number of cells, re-sampling to different cadences) we generally find that the \ac{mcoa} method has a higher success rate in reconstructing the intrinsic \ac{evpa} curve correctly.

We find that the usual \ac{mcoa} method is more reliable.  The \ac{mcor} method has not proven useful, so we adopt the \ac{mcoa} method for the rest of this paper.  Note, however, that if a different rotation model is proposed, then these two well-motivated methods should be tested and compared before choosing which to apply.

\subsection{The Results of season 1-3 and season 4 after adjustment}
\label{sec:results}

\subsubsection{Reliability of identified rotations}
\label{sec:rotationreliability}

Using the criteria described above,  after adjusting the \ac{evpa} curves we find 43~rotations during seasons~1--3 in the full sample. The season~4 sample is a subset of 28 objects taken from the full sample (see Table \ref{tab:sources}). Amongst these 28 objects we identify 30~rotations in seasons~1--3, and 9~rotations in season~4. The identified rotation periods of the season~4 sample are shown highlighted in  \cref{app:rotations}.

\begin{figure}
  \centering
  \includegraphics[width=\columnwidth]{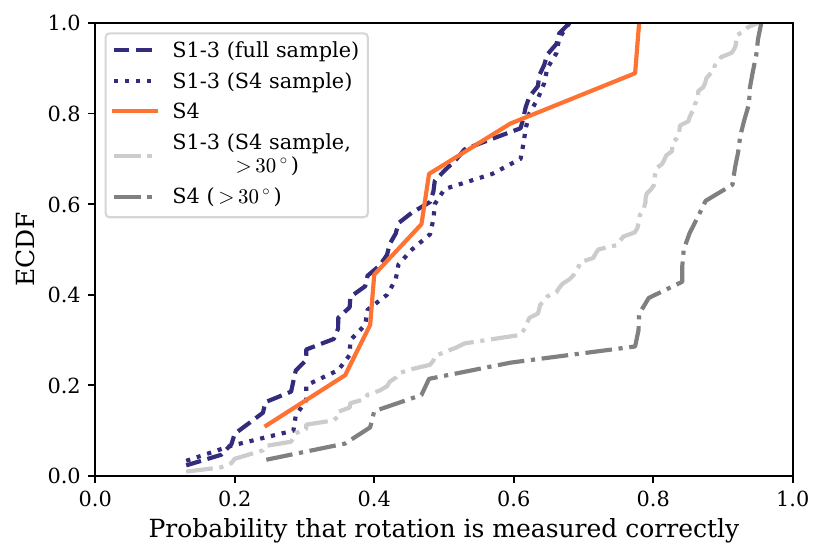}
  \caption{
    \ac{ecdf} of the probability that a rotation was measured correctly for different samples of rotations identified in the RoboPol sample. Purple, dashed line: rotations identified in seasons~1--3 of the full sample. Purple, dotted line: rotations identified in seasons~1--3 of the season~4 sample. Orange, solid line: rotations from season~4. Light grey, dash-dotted line: rotations identified in seasons~1--3 of the season~4 sample, including rotations as short as $30^\circ$. Dark grey, dash-dotted line: rotations identified in seasons~4, including rotations as short as $30^\circ$.}
  \label{fig:rotationreliability}
\end{figure}

As described in \cref{app:probability}, \cref{eq:probgood} can be used to estimate the probability that a measured \ac{evpa} change, $\Delta\chi_\mathrm{wrap}$, between two consecutive data points with time separation $\Delta t$, correctly represents the intrinsic \ac{evpa} change, i.e. that its absolute value did not intrinsically exceed $90^\circ$ and was thus not affected by the $180^\circ$~ambiguity.
The probability that an \ac{evpa} rotation event was measured correctly is then the product of such probabilities for all consecutive data pairs.
We note that successive \ac{evpa} measurements are not independent random variables, but are related through a -- currently unknown -- physical process. Through this process the distribution of an \ac{evpa} change is constrained by the preceding change(s). As this process is currently unknown, we here treat the measurements as independent random variables. This is the most conservative approach, as further constraints on the probability density function of \ac{evpa} changes would increase the estimated probability that a rotation was measured correctly.
For each identified rotation we calculate  the probability that it was measured correctly.
\Cref{fig:rotationreliability} shows the \ac{ecdf} of the resulting probabilities for rotations of amplitude $>90^\circ$ identified in seasons~1--3 (dashed and dotted lines) and in season~4 (solid line).
We find that $\sim65\%$ of the identified rotations are at least as likely to be measured incorrectly as they are to be measured correctly. Therefore, although a small fraction of \ac{evpa} changes ($\sim 11\%$ for seasons~1--3) are expected to be affected by the $180^\circ$~ambiguity, the probability that a \emph{rotation event} is affected (by having at least one affected consecutive measurement pair) is substantial. 
The inclusion of rotations with smaller amplitudes $>30^\circ$, which are less fast, adds rotations with significantly higher probability that they were not affected by the $180^\circ$~ambiguity (dash-dotted grey lines).

\subsubsection{Rotation rates}
\label{sec:rates}

For each observed rotation event, we measure its amplitude, duration, and rate.
The rotation \emph{amplitude} is the  absolute value of the difference in \ac{evpa} between the last and the first data point.
The rotation \emph{duration} is the time interval between the first and last observations of the event.
We estimate the average rotation \emph{rate} by dividing the amplitude by the duration.

\begin{figure}
  \centering
  \includegraphics[width=\columnwidth]{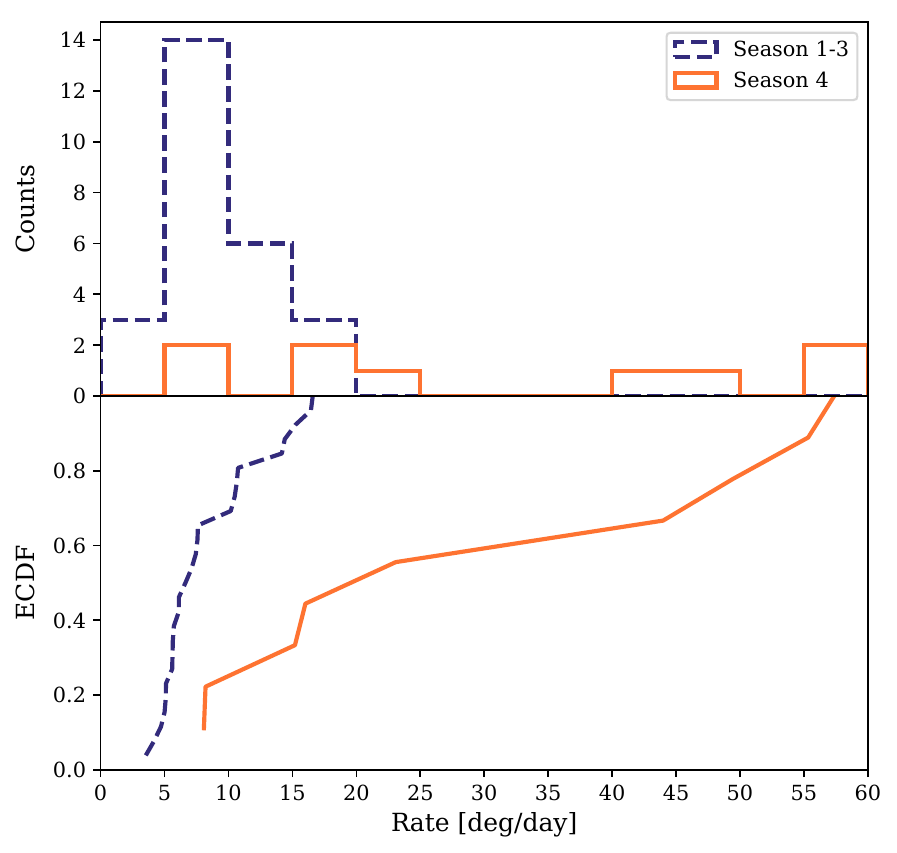}
  \caption{
    Histogram (upper panel) and \ac{ecdf} (lower panel) of rotation rates of identified rotation candidates in seasons~1--3 (purple) and season~4 (orange) in the RoboPol sample.}
  \label{fig:histrate}
\end{figure}

In comparing the rotation rates in seasons~1--3 and season~4, we consider only the common sources, i.e. the season~4 subsample, and we exclude four~rotations from seasons~1--3 whose duration exceeds the median observing period of season~4, which we would not have been able to detect in the short period of season~4.
The rotation rates are shown in \cref{fig:histrate} and corresponding statistics are listed in \cref{tab:rotrate_stats}.
Rotations identified in the season~4 data rotate faster, on average than rotations identified in seasons~1--3.
In fact, the majority of rotations in season~4 rotate faster than the fastest one detected in seasons~1--3.
With a cadence of 7~days the detectable rotation rates are limited by the ambiguity to $<90^\circ / 7~\mathrm{days} \approx 12.8^\circ/\mathrm{day}$.
Thus, the majority of rotations found in season~4 could not have been detected with the average cadence of seasons~1--3.

We also find that the majority of rotations identified in seasons~1--3 are slower than the slowest one detected in season~4.
We discuss this lack of slow rotations in the daily sampled data in \cref{app:extrinsic}.

\begin{table}
    \caption{
        Statistics of the distributions of rotation rates for the rotations identified in the RoboPol sample during seasons~1--3 compared to season~4.
        The corresponding uncertainties are estimated with a bootstrap method; in 100~iterations a random fraction of rotation events is selected and the analysis is repeated on this selection; for each measured property the uncertainty is given by the standard deviation over all bootstrap iterations.}
    \label{tab:rotrate_stats}
    \centering
    \begin{tabular}{l r r r r}
        \toprule
        & min & median & mean & max \\
        & [deg/day] & [deg/day] & [deg/day] & [deg/day]\\
        \midrule
        Season 1--3: & $  3.5 \pm   0.5$ & $  6.9 \pm   0.8$ & $  8.4 \pm   0.6$ & $ 16.6 \pm   0.5$ \\
        Season 4: & $  8.1 \pm   2.2$ & $ 23.1 \pm  11.3$ & $ 30.7 \pm   5.0$ & $ 57.4 \pm   1.8$ \\
        \bottomrule
    \end{tabular}
\end{table}{}

\subsection{The effect of a faster cadence}
\label{sec:faster}

\begin{figure}
  \centering
  \includegraphics[width=\columnwidth]{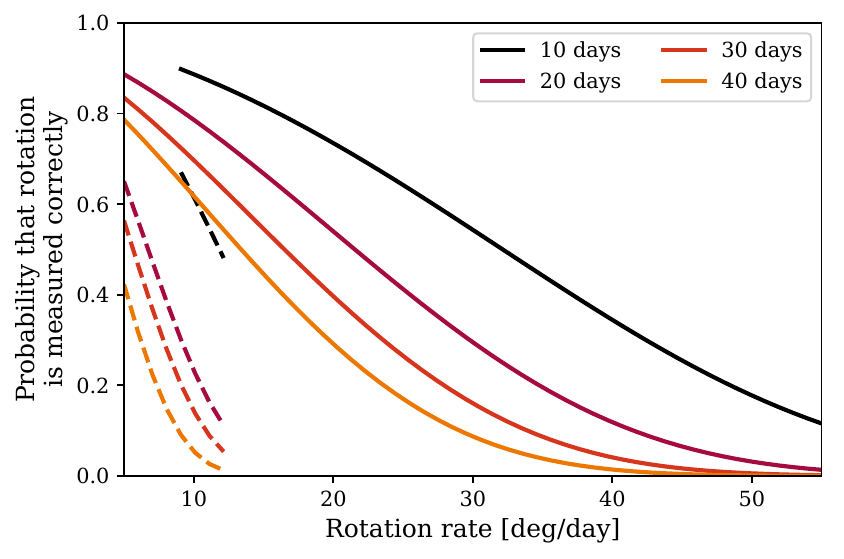}
  \caption{
    Probability that a detected rotation is measured correctly for different measured rotation rates, assuming the \ac{evpa} distribution model derived from the RoboPol data. Different colours correspond to different rotation durations as stated in the legend. Solid lines correspond to daily cadence -- the cadence of RoboPol season~4 --, dashed lines to weekly cadence -- the mean cadence of RoboPol seasons~1--3.}
  \label{fig:rotconfidence}
\end{figure}

Although it is obvious that faster cadences must lead to an improvement in the reliable detection of more rotations, the magnitude of the effect is not so obvious. To demonstrate the magnitude of the effect, we assume that we detect rotations with a constant rotation rate and a certain duration with a given, constant cadence of observations. We can use the formalism described in \cref{sec:rotationreliability} to estimate the probability that the detected rotation correctly represents the intrinsic variability. This probability represents the confidence we have in a detected rotation.
\Cref{fig:rotconfidence} shows the confidence for different combinations of rotation rates and durations in the ranges that we found in the RoboPol data. The confidence is plotted for the median cadence of season~4 and of seasons~1--3. We note that, by definition, combinations of rate and duration that lead to a rotation amplitude lower than $90^\circ$ are not identified as rotations in this study (except in the single instance where we use the $30^\circ$ lower limit). Rotation rates that lead to an \ac{evpa} change larger than $90^\circ$ cannot be detected due to the ambiguity, this limits the detected rotation rates for a given cadence in this study.
In addition, we do not require that the rotations are sampled with at least four data points. Otherwise, rotations with a duration $<21$~days would not be detectable with weekly cadence.
The comparison of the dashed and solid lines (of the same colour) in \cref{fig:rotconfidence} demonstrates how strongly the confidence in detected rotations increases with faster observing cadence.
Furthermore, \cref{fig:rotconfidence} allows us to estimate the ranges of rotation rates and durations that would be detectable with daily sampling in a future monitoring program for an a~priori defined confidence limit.

\section{Discussion}
\label{sec:discussion}

The daily sampled season~4 data reveal a number of significantly faster rotations than were identified in season 1--3. Thus we have clearly missed a number of rapid rotation events in seasons~1--3 due to the 7-day cadence.
We would have detected significantly more and significantly faster rotations in seasons~1--3 of the RoboPol observations with a 1-day cadence.
We showed that the detected rotations in the RoboPol sample cover a large range of rotation rates up to $57^\circ/\mathrm{day}$.
This is not a physical upper limit, but a result of the limited cadence and observation duration.
Significantly faster rotations have been described in the literature, as we discuss further below.
During RoboPol season~4, which lasted less than two~months, only nine~rotations were identified using weekly cadence.
This number of events is not sufficient to constrain the distribution of rotation rates.

Models of rotations should take the large range of measured rotation rates into account and need to be able to produce rotations with a variety of amplitudes, durations, and rates.
We note, however, that the distributions shown here depend on the specific definition of a rotation event, the cadence of the observations, and the sample.
The same set of criteria need to be used for the comparison of data and models.

With an independent method we confirmed the results of \citet{blinov_2015,kiehlmann_2017} that the \ac{evpa} progression is not consistent with a simple random walk.
This result challenges the turbulence based model of \citet{2014ApJ...780...87M}.
The method used to test the simple random walk model here, can be applied to test any model that aims to reproduce the typical \ac{evpa} variability in blazars.

With the method described in \cref{sec:ambiguity} we have for the first time determined how strongly the \ac{evpa} curves of blazars are affected by the ambiguity for different separations.
We found that the ambiguity affects data on all tested separations from 1--30~days.
Sampled with 7~days cadence -- the average cadence of RoboPol observations during seasons~1--3 -- we expect 11\% of \ac{evpa} changes to exceed $90^\circ$, leading to false estimates of the \ac{evpa} distribution.
A daily cadence leads to a significant improvement, since in this case only 4\% of the data are expected to exceed $90^\circ$. Our method thus enables us to estimate our confidence in the identified rotations. It shows that at least a 1-day cadence is needed in such studies. We identified rotations in four seasons of RoboPol data and estimated that about 60\% of the rotations are more likely to be measured incorrectly than correctly due to the $180^\circ$ ambiguity.

We caution the reader that these results are specific for the selected sample and may only be extended to sources that satisfy the same selection criteria.
In particular the specific combination of \textit{Fermi}-\ac{lat}-detected and -nondetected may have an effect on the quantitative results.
However, a separate study of these sub-samples is beyond the scope of this work.
Furthermore, we note that the estimates of the expected fraction of \ac{evpa} changes exceeding $90^\circ$ and the estimates of the reliability of identified rotations depend on the model assumption that the intrinsic \ac{evpa} changes follow a log-normal model. Future observations -- in particular using faster cadence -- and physical model simulations may allow us to test this assumption and/or better select and constrain the distribution model.

In many sources it is the periods of fastest \ac{evpa} changes that lead to their identification as a rotation.
\Cref{sec:ambiguity} shows that even daily observations -- as in the case of RoboPol season~4 -- are not sufficient to avoid the $180^\circ$~ambiguity in the fastest varying sources.
\citet{2020ApJ...902...61L} recently reported an \ac{evpa} rotation of $230^\circ$ in 2~days in 3C~454.3.
If the measured rotation correctly represents the intrinsic \ac{evpa} progression, the rotation rate exceeds the rate of the fastest rotation detected in the RoboPol data by a factor of~2.
The data used by \citet{2020ApJ...902...61L} included RoboPol and other instruments.
Multiple instruments gave a cadence faster than 1 day, as is clearly required to measure such fast variability. The detected rotation included a large jump close to $90^\circ$, showing that even in this case the cadence  was barely adequate.
The fastest \ac{evpa} rotation so far was reported by \citet{2018A&A...619A..45M} in S5~0716+714 at MJD~57044-57052, showing a $\sim~400^\circ$ change of the \ac{evpa} in less than one day, corresponding to an average rotation rate of $400^\circ/\mathrm{day}$ with an extremely fast onset of $~300^\circ$ in $\sim3.6$~hours, corresponding to a peak rotation rate of $\sim2000^\circ/\mathrm{day}$.
A rotation at this rate requires a cadence of at least one observation every 140~minutes to avoid under-sampling.
Thus, to track the fastest \ac{evpa} changes correctly -- assuming this particular event was measured correctly -- continuous monitoring with multiple telescopes around the globe is necessary.
Our model suggests that a rotation this fast or faster at the separation of hours is an unlikely event with a probability of $1.3\times10^{-4}$; however our model was not informed by data sensitive to such fast variations. 
A campaign of the same scale as RoboPol but with significantly better cadence is needed to study the distribution of such rapid rotations.

\section{Conclusions}
\label{sec:conclusions}

We used three seasons of RoboPol optical polarization monitoring data sampled with approximately weekly cadence and one season of daily observations to identify \ac{evpa} rotations in a statistical sample of blazars.
We showed that the rotation speeds cover a wide range up to $57^\circ/\mathrm{day}$.
The two different cadences allowed us to test the effects of cadence on the identification of rotations.
Due to the $180^\circ$ ambiguity the fastest rotations detected require daily or faster cadence and many fast rotations must have passed undetected in the weekly sampled RoboPol data.
Furthermore, the definition of a rotation event limits which periods are detected as a rotation.
The definition explicitly introduced for the weekly sampled data, may need to be revised for better sampled data.

We studied how strongly the \ac{evpa} varies on different time scales and showed that \ac{evpa} changes may exceed $90^\circ$ on all tested time scales $>1$~day.
Therefore, the \ac{evpa} measurements may be affected by the $180^\circ$ ambiguity on all time scales $>1$~day. Shorter time scales could not be tested with the RoboPol data.
We introduced a procedure that allowed us to estimate the fraction of data that is expected to exceed $90^\circ$ on different time scales.
We estimated that $11\%$ of the RoboPol data sampled with weekly cadence and the majority of the identified rotations are likely affected by the ambiguity.
Daily cadence leads to a significant improvement, as only $4\%$ of the data are affected.
We caution that these results are specific for the studied sample and may differ for other samples of blazars.

Season~4 of the RoboPol program lasted only about 45~days and did not provide the long-term monitoring data necessary for a revision of the definition of \ac{evpa} rotation events and to establish a large set of reliable rotation events for model testing.
We clearly need optical monitoring programs of the same scope as RoboPol, but with a cadence significantly faster than 1~day, which requires multiple observing sites. For this reason we are now planning a second RoboPol instrument for deployment at a substantially different longitude.

\section*{Acknowledgements}

The authors thank the anonymous referee for the positive and constructive response that helped to improve this manuscript.
The authors acknowledge the contributions of O.~G.~King, A.~Kus and E.~Pazderski to the RoboPol project.
The RoboPol project is a collaboration between Caltech in the USA, Max-Planck-Institute for Radio Astronomy in Germany, Toru\'n Centre for Astronomy in Poland, the University of Crete/FORTH in Greece, and IUCAA in India.
This research was supported in part by NASA grant NNX11A043G and NSF grant AST-1109911, and by the Polish National Science Centre, grant numbers 2011/01/B/ST9/04618 and 2017/25/B/ST9/02805.
D.B., C.C., S.K., N.M., R.S., and K.T. acknowledge support from the European Research Council under the European Union's Horizon 2020 research and innovation programme, grant agreement No771282.
V.P. acknowledges support from the Foundation of Research and Technology - Hellas Synergy Grants Program through project MagMASim, jointly implemented by the Institute of Astrophysics and the Institute of Applied and Computational Mathematics and by the Hellenic Foundation for Research and Innovation (H.F.R.I.) under the ``First Call for H.F.R.I. Research Projects to support Faculty members and Researchers and the procurement of high-cost research equipment grant'' (Project 1552 CIRCE).
A.N.R., G.V.P., and A.C.S.R. acknowledge support from the National Science Foundation, under grant number AST-1611547.
G.V.P. acknowledges support by NASA through the NASA Hubble Fellowship grant \#~HST-HF2-51444.001-A awarded by the SpaceTelescope Science Institute, which is operated by the Association of Universities for Research in Astronomy, Incorporated, under NASA contract NAS5-26555.
T.J.P. acknowledges support from NASA grant NNX16AR31G.
T.H. was supported by the Academy of Finland projects 317383, 320085, and 322535.
A.N.R acknowledges support through a grant from the Infosys Foundation.
This research made use of Stan, \url{https://mc-stan.org/}, through the PyStan interface, \url{https://pystan.readthedocs.io/}, Numpy \citep{numpy}, SciPy \citep{scipy}, StatsModels \citep{statsmodels}, Matplotlib \citep{matplotlib}, and CMasher \citep{cmasher}.

\section*{Data Availability}

The data underlying this article are available in ``RoboPol: AGN polarimetric monitoring data'', at \url{https://doi.org/10.7910/DVN/IMQKSE}.

\bibliographystyle{mnras}
\bibliography{references}

\appendix

\section{Model of EVPA changes}
\label{app:model}

The first part of this appendix section describes how we estimate the distribution of intrinsic \ac{evpa} changes from the distribution of measured, wrapped \ac{evpa} changes for different cadences. The results are discussed in \cref{sec:ambiguity}.
The second part describes how we use the model to estimate the probability that a measurement was affected by the $180^\circ$~ambiguity.

\subsection{Model description}

We use the following empirical model to describe the distribution of wrapped \ac{evpa} changes. 
As discussed in \cref{sec:model}, the distributions of adjusted \ac{evpa} changes on different time scales resemble a log-normal distribution with changing distribution parameters.
Based on this observation, we model the \emph{intrinsic distribution} of the absolute \ac{evpa} change $|\Delta\chi|_{\Delta t}$ at a time scale $\Delta t$ as a log-normal distribution:
\begin{align}
    \mathrm{PDF}_\mathrm{intr}(x) = \mathcal{LN}(x; \mu, \sigma) = \frac{1}{\sqrt{2\pi}\sigma x}\exp\left(-\frac{\left(\ln(x) - \mu\right)^2}{2\sigma^2}\right),
\end{align}
with  the mean, $\mu$, and the standard deviation, $\sigma$ of the variable's natural logarithm, and  $x = |\Delta\chi|_{\Delta t}$.
We point out that the choice of a log-normal distribution is an assumption about the distribution of the intrinsic \ac{evpa} changes.
Physical model simulations of the optical polarized emission of blazars may help to select a physically motivated distribution model in the future.
The intrinsic distribution cannot be directly measured, because we can only measure differences up to $90^\circ$ between two consecutive data points, due to the $180^\circ$~ambiguity.

\begin{figure}
    \centering
    \includegraphics[width=\columnwidth]{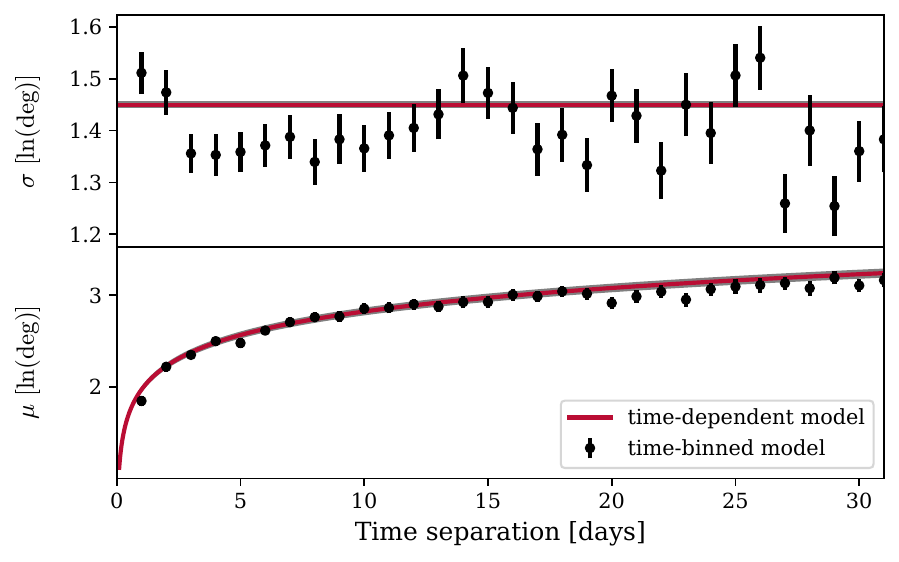}
    \caption{
        Model parameters derived for different time scales. Data points: Results of the time-binned data using a bin width of 1~day. Error bars indicate the $1\sigma$-credible intervals. Red line: Result of the time-dependent model. The thin, grey region around the red lines indicates the $1\sigma$-credible interval.}
    \label{fig:params}
\end{figure}

\begin{figure*}
	\begin{minipage}{0.5\textwidth}
	   \centering
	   \includegraphics[width=1\textwidth]{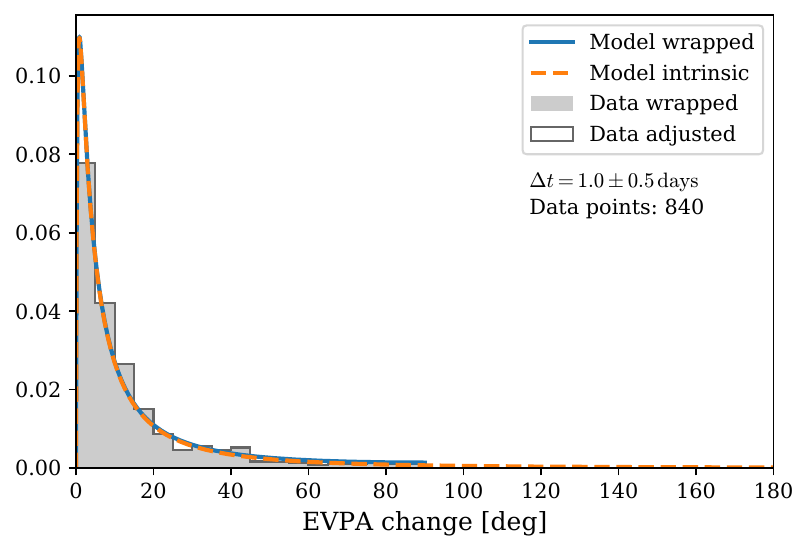}
	\end{minipage}%
	\begin{minipage}{0.5\textwidth}
	   \centering
	   \includegraphics[width=1\textwidth]{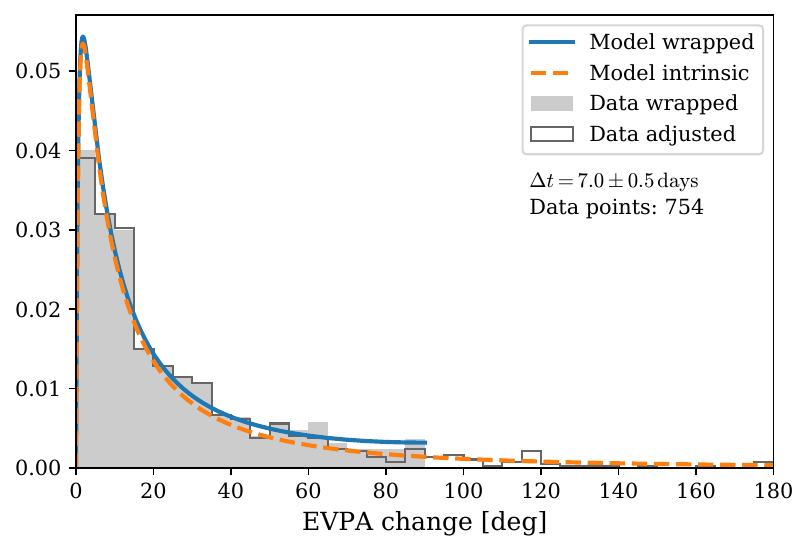}
	\end{minipage}
	\hfill
	\begin{minipage}{0.5\textwidth}
	   \centering
	   \includegraphics[width=1\textwidth]{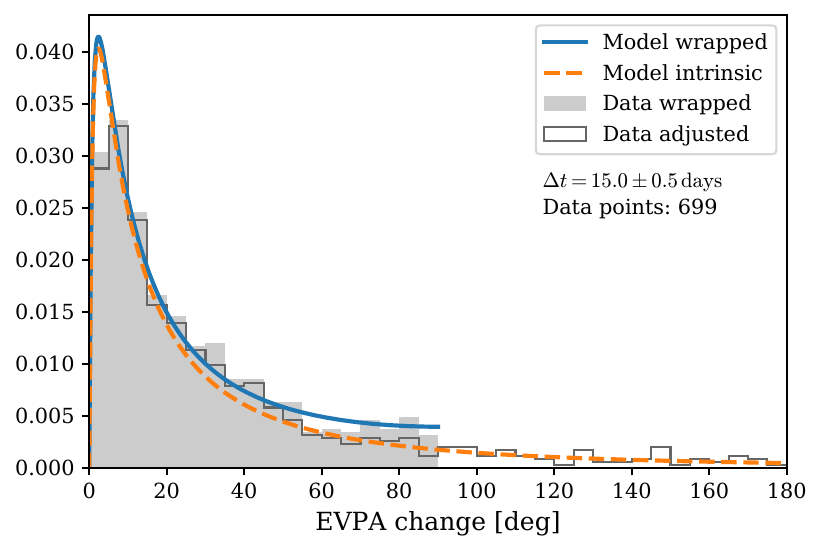}
	\end{minipage}%
	\begin{minipage}{0.5\textwidth}
	   \centering
	   \includegraphics[width=1\textwidth]{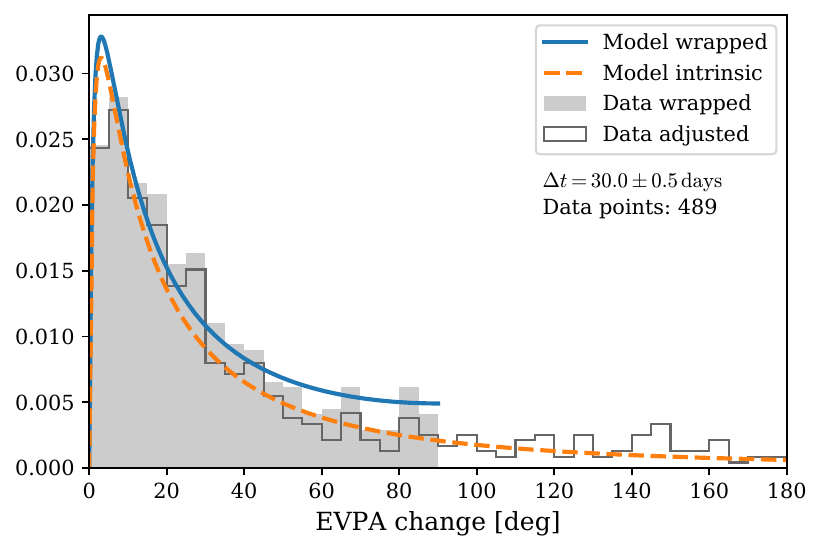}
	\end{minipage}
    \caption{Examples of $\mathrm{PDF}_\mathrm{meas}$ (blue solid line) and $\mathrm{PDF}_\mathrm{intr}$ (orange dashed line) for the best-fit parameters $\sigma$ and $\mu$, for different timescales. The best-fit parameters were obtained through fitting $\mathrm{PDF}_\mathrm{meas}$ to the distribution of wrapped \ac{evpa} changes (grey area) measured from the RoboPol data. Distributions of \ac{evpa} changes from the adjusted RoboPol \ac{evpa} curves (grey outline) are shown in comparision to $\mathrm{PDF}_\mathrm{intr}$. See~\cref{sec:ambiguity} for the definition of wrapped and adjusted \ac{evpa} changes. Each panel corresponds to a different time scale, indicated in the legend. A timescale of $\Delta t \sim 1$~day (top left) corresponds to the cadence of RoboPol season~4, and 7~days (top right) to the average cadence of seasons~1--3. The timescale of 30~days (bottom right) is the longest timescale considered in our analysis.}
    \label{fig:modelfits}
\end{figure*}

The \emph{wrapped distribution} $\mathrm{PDF}_\mathrm{wrap}$ can be described by a modified version of the log-normal distribution:
\begin{align}
    \mathrm{PDF}_\mathrm{meas}(x) =
    \begin{cases}
        \mathcal{LN}_\mathrm{wrap}(x; \mu, \sigma) \text{ for } x\in[0^\circ,90^\circ] \\
        0 \text{ otherwise}
    \end{cases},
    \label{eq:pdfintrinsic}
\end{align}
which can be derived from $\mathrm{PDF}_\mathrm{intr}$ as follows. 
If the \ac{evpa} intrinsically changes by e.g. $20^\circ$, we measure $|\Delta\chi|=20^\circ$. If intrinsically it \ac{evpa} changes $160^\circ$, we would measure it as $\Delta\chi=-20^\circ$, i.e. $|\Delta\chi|=20^\circ$. For an intrinsic change of $200^\circ$, we would measure $\Delta\chi=20^\circ$, i.e. $|\Delta\chi|=20^\circ$.
The probability of measuring a value $x\in[0, 90]$ is:
\begin{align}
    \mathcal{LN}_\mathrm{wrap}(x; \mu, \sigma)
    &= \mathcal{LN}(x; \mu, \sigma) \nonumber\\
    &+\mathcal{LN}(180^\circ-x; \mu, \sigma) \nonumber\\
    &+ \mathcal{LN}(180^\circ+x; \mu, \sigma) \nonumber\\
    &+ \mathcal{LN}(360^\circ-x; \mu, \sigma) \nonumber\\
    &+ \cdots
    \label{eq:pdfmeasured}
\end{align}
The full expression can be written as:
\begin{align}
    \mathcal{LN}_\mathrm{wrap}(x; \mu, \sigma) = \lim\limits_{N\to\infty} \sum\limits_{n=0}^N \mathcal{LN}(x_n; \mu, \sigma)
    \label{eq:lnmod}
\end{align}
with
\begin{align}
    x_n = 90 (n + m(n)) + (-1)^n x,
    \label{eq:xn}
\end{align}
where $m(n)$ is a function that is 0 for even numbers and 1 for odd numbers, for which we choose $m(n) = \sin^2(\frac{n\pi}{2})$.

One may think of this modified distribution as such: We print the lognormal distribution on paper, every $90^\circ$ on the x-axis we wrap the paper parallel to the y-axis, lastly we sum up all probability density values for each x-value between 0 and $90^\circ$.\footnote{The formalism described in \crefrange{eq:pdfmeasured}{eq:xn} fails at $0^\circ$ and $90^\circ$, because only every second term should be added.
However, since we never measure exactly $0^\circ$ or $90^\circ$, and the discontinuity resulting from this feature does not affect our results in any way, we have retained and implemented this simple version of $\mathcal{LN}_\mathrm{wrap}$ described above.
}


Rather than $N\to\infty$ for the implementation of \cref{eq:lnmod} we have to choose an $N$ sufficiently large.
We kept this a modifiable parameter that we finally choose large enough that larger values do not show a significant impact on the final results.
For the final model fitting we chose $N=10$ and found that the inferred parameters do not differ significantly if we use $N=6$.

\textbf{Time-binned model:}
We implement the following model in \verb|pystan|, using uniform distributions, $\mathcal{U}(0, 10^2)$, as diffuse priors for the model parameters $\mu,\sigma$:
\begin{align}
    Y_i &\sim \mathcal{LN}_\mathrm{wrap}(\mu, \sigma^2) \\
    \mu &\sim \mathcal{U}(0, 10^2) \\
    \sigma &\sim \mathcal{U}(0, 10^2) \\
    i &= 1, \dots, M
\end{align}
where $Y_i$ are the measured \ac{evpa} changes and $M$ is the number of data points.


We bin the wrapped \ac{evpa} changes according to their  corresponding time scales, using a bin width of 1~day, and we use the binned data to infer the optimal parameters, $\sigma$ and $\mu$, of the model $\mathrm{PDF}_\mathrm{wrap}$ described above, on different timescales.
\cref{fig:params} shows the model parameters for different separations.
Parameter $\sigma$ shows no clear dependence on the separation and the differences are sufficiently small -- considering the credible intervals -- that we may assume it constant. On the other hand, parameter 
$\mu$ does show a dependence on the timescale that can be expressed as a linear function of the logarithm of the separation, $\mu(\Delta t) = \beta_0 + \beta_1 \log(\Delta t)$.

\textbf{Timescale-dependent model:}
We include this dependence in our Bayesian modelling frame work and fit the entire data of time differences and wrapped \ac{evpa} changes with a single model:
\begin{align}
    Y_i &\sim \mathcal{LN}_\mathrm{wrap}(\mu_i, \sigma^2) \\
    \mu_i &\equiv \beta_0 + \beta_1 \log(x_i)  \\
    \sigma &\sim \mathcal{U}(0, 10^2) \\
    \beta_j &\sim \mathcal{U}(0, 10^2) \\
    j &= 1, 2  \\
    i &= 1, \dots, M
\end{align}
where $x_i=\Delta t_i$ are the timescales corresponding to the wrapped \ac{evpa} changes $Y_i=|\Delta\chi_\mathrm{wrapped}|$.
We use diffuse priors for the three model parameters, $\sigma, \beta_0, \beta_1$.
The fit results are discussed in \cref{sec:ambiguity}.
\cref{fig:modelfits} shows four examples of different separation bins, for which the data is compared to the model with the best-fit parameters. Note that the discrepancy between model and data at very small values of the \ac{evpa} change (i.e. the model peak close to zero that is not reflected in the observed histogram) is an expected effect of our finite measurement uncertainty in the \ac{evpa}, which has not been explicitly implemented in our treatment. Specifically, if an \ac{evpa} change $\Delta \chi$ was consistent with zero within uncertainties, we recorded its actual measured value rather than zero. This results in a "flattening" of the small-\ac{evpa} peak that the model (correctly) exhibits.

\subsection{Estimated probability of under-sampled measurement}
\label{app:probability}

Let us assume we measure $\Delta\chi_\mathrm{wrap} \in [0^\circ, 90^\circ]$ and the intrinsic \ac{evpa} change equals the measured one, $\Delta\chi_\mathrm{intr} = \Delta\chi_\mathrm{wrap}$, i.e. was measured correctly.
Then, we can express the joint probability density, $P(\Delta\chi_\mathrm{intr}{=}\Delta\chi_\mathrm{wrap} \cap \Delta\chi_\mathrm{wrap})$, through the intrinsic distribution in \cref{eq:pdfintrinsic} for any given $\Delta\chi_\mathrm{wrap} \in [0^\circ, 90^\circ]$.
The probability that we measure the intrinsic \ac{evpa} change correctly, given a certain measurement $\Delta\chi_\mathrm{wrap}$ is:
\begin{align}
    P(\Delta\chi_\mathrm{intr}=\Delta\chi_\mathrm{wrap} | \Delta\chi_\mathrm{wrap})
    &= \frac{P(\Delta\chi_\mathrm{intr}{=}\Delta\chi_\mathrm{wrap} \cap \Delta\chi_\mathrm{wrap})}{P(\Delta\chi_\mathrm{wrap})} \nonumber\\
    &= \frac{\mathrm{PDF}_\mathrm{intr}(\Delta\chi_\mathrm{wrap};\mu, \sigma(\Delta t))}{\mathrm{PDF}_\mathrm{meas}(\Delta\chi_\mathrm{wrap};\mu, \sigma(\Delta t))}
    \label{eq:probgood}
\end{align}
The probability that the intrinsic \ac{evpa} exceeded $90^\circ$, and thus was measured incorrectly, is:
\begin{align}
    P(\Delta\chi_\mathrm{intr}>90^\circ | \Delta\chi_\mathrm{wrap})
    = 1 - \frac{\mathrm{PDF}_\mathrm{intr}(\Delta\chi_\mathrm{wrap};\mu, \sigma(\Delta t))}{\mathrm{PDF}_\mathrm{meas}(\Delta\chi_\mathrm{wrap};\mu, \sigma(\Delta t))}
    \label{eq:probbad}
\end{align}
With the model parameters, $\mu, \sigma(\Delta t)$, estimated from the model fit discussed in the previous section, we can estimate the probability that a measured \ac{evpa} change, $\Delta\chi_\mathrm{wrap}$, does (not) represent the intrinsic \ac{evpa} changes with \cref{eq:probgood,eq:probbad}, for any given data pair $(\Delta t, \Delta\chi_\mathrm{wrap})$.

\section{Extrinsic Factors in Identifying Rotations}
\label{app:extrinsic}

In the following we test how the extrinsic factors of cadence, length of observing period and smoothness  affect the identification of rotations and the analysis of the rotation parameters.

\subsection{Examples of sampling effects}
\label{sec:example}

\begin{figure}
  \centering
  \includegraphics[width=\columnwidth]{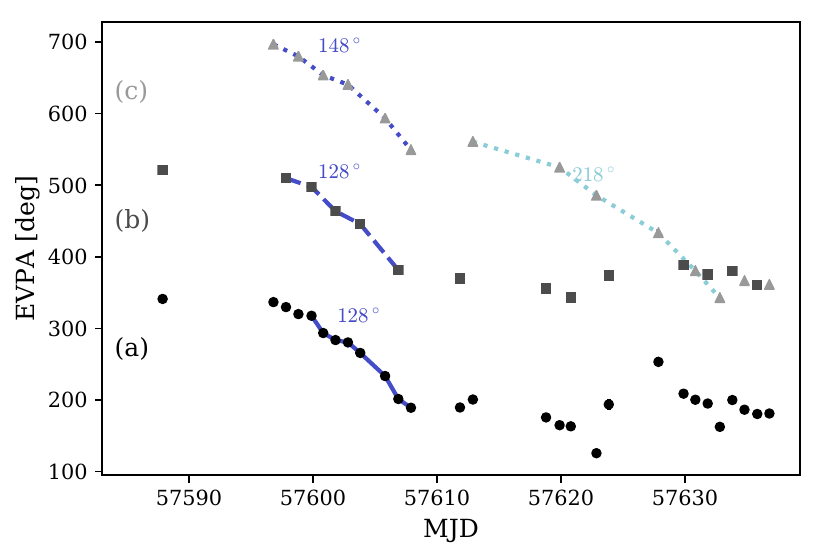}
  \caption{
    Illustration of the effects of different cadences on the identification of \ac{evpa} rotations. (a) Season~4 data of RBPLJ1635+3808 (black circles) and corresponding rotation (solid line). (b) Every second data point of the original data (dark grey squares) and corresponding rotation (dashed line). (c) As in (b), but using the data points that were omitted in (b) (light grey triangles). Rotations in this case are shown with the dotted lines. Each \ac{evpa} curve was separately adjusted for the $180^\circ$~ambiguity.}
  \label{fig:samplingeffects}
\end{figure}

In \cref{fig:samplingeffects}, we illustrate the type of cadence effects that may affect \ac{evpa} curve analyses, using the densely-sampled season 4 RoboPol data for source RBPLJ1635+3808. 
The complete \ac{evpa} data are plotted using black dots and are marked as~(a).
We also show two realisations of the same data with slower cadence, by removing every second data point, starting either with removing the second point~(b) or with removing the first point~(c).
For clarity we have shifted the three curves by $180^\circ$.
Realizations~(b) and~(c) were individually adjusted for the $180^\circ$~ambiguity (c.f. \cref{sec:ambiguity}) after the removal of data points from the original \ac{evpa} curve.

In the full \ac{evpa} curve (a) we identify one rotation with amplitude~$128^\circ$ in the first half of season 4.
In realization~(b) we also identify one rotation of the same amplitude in the first half of season 3, but slightly shifted in time.
In realization~(c) we identify a longer rotation of~$148^\circ$ in the first half of season 4 as well as a longer rotation in the second half of the season.
The rotations in the first half of season 4 in both under-sampled realizations include data points that were not considered part of the rotation in the original data (either before the beginning or after the end of the rotation seen in the full data).
The reason is that the data that are more densely sampled reveal short-timescale \ac{evpa} changes that violates our definition of a smooth rotation.

After MJD~57610 the full dataset shows \ac{evpa} changes with changing directions.
Realization~(b) appears more stable in comparison.
Realization~(c), however, shows a rotation of~$218^\circ$, because the removal of one critical data point led to a differently adjusted \ac{evpa} curve.
This example demonstrates how a slower cadence can result in an apparently larger range of \ac{evpa} changes.

These examples indicate two potential problems in \ac{evpa} rotation measurements:

\begin{enumerate}
    \item \ac{evpa} changes on short time-scales may be strong enough for the candidate rotation to be rejected due to our smoothness criterion (c.f. \cref{sec:samplingdependence}). In such cases long rotations may only be identified in under-sampled data.
    \item Sparse sampling of fast \ac{evpa} changes can critically affect the identification of rotation periods.
\end{enumerate}


\subsection{Effect of length of observing season, cadence, and smoothness on derived rotation parameters}
\label{sec:rotpar}

In \cref{sec:rates} we saw that season~4 shows significantly faster rotations than seasons~1--3, because the cadence of seasons~1--3 was too slow to detect such fast rotations.

\begin{figure}
  \centering
  \includegraphics[width=\columnwidth]{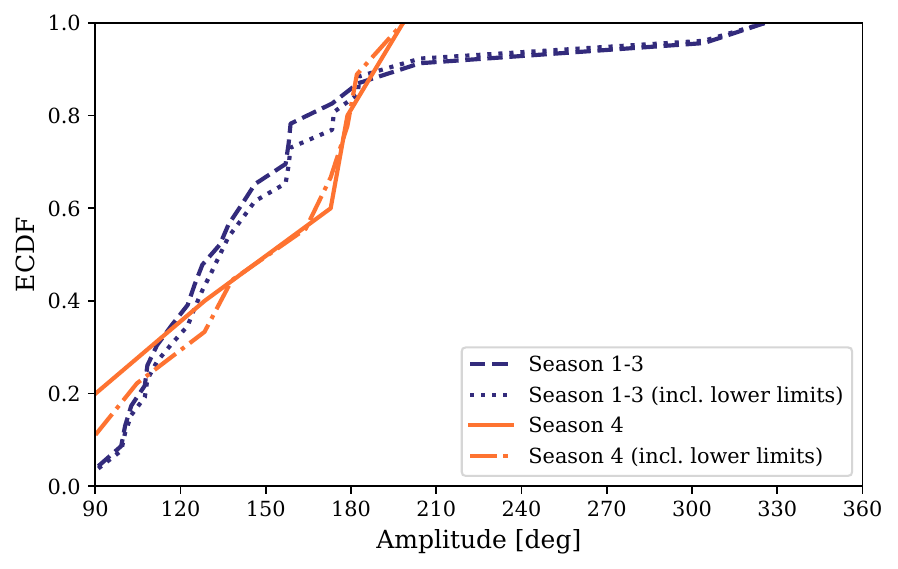}
  \caption{
    \ac{ecdf} of amplitudes of identified rotation candidates in seasons~1--3 (purple) and season~4 (red) of the RoboPol data. Solid lines: distributions excluding lower limits. Dashed lines: distributions including lower limits.}
  \label{fig:histamplitude}
\end{figure}

\begin{figure}
  \centering
  \includegraphics[width=\columnwidth]{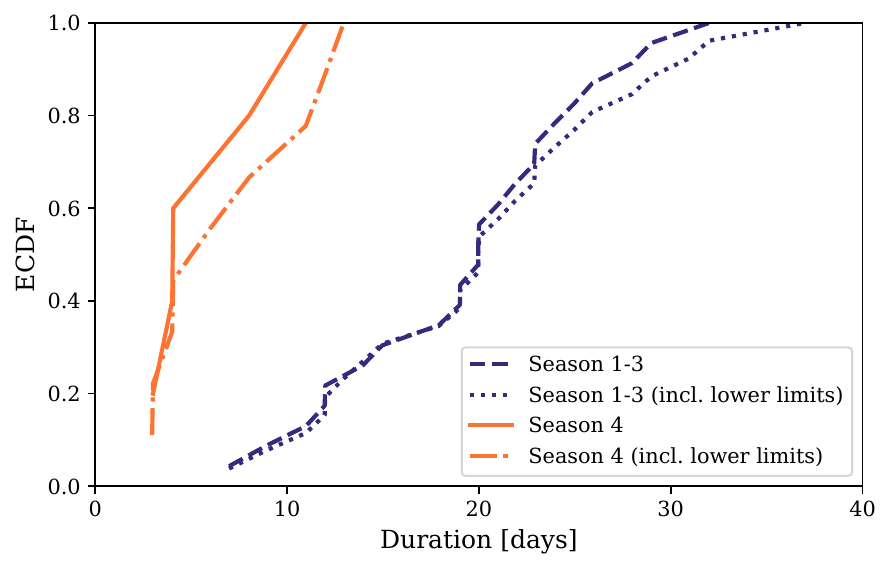}
  \caption{
    \ac{ecdf} of durations of identified rotation candidates in seasons~1--3 (purple) and season~4 (orange) of the RoboPol data. Solid lines: distributions excluding lower limits. Dashed lines: distributions including lower limits.}
  \label{fig:histduration}
\end{figure}

Here, we discuss the apparent lack of slow rotations in season~4.
\cref{fig:histrate} shows that $\sim60\%$ of the rotations detected in seasons~1--3 are slower than the slowest rotation detected in season~4.
The average rotation rates are calculated from the amplitude divided by the duration.
\cref{fig:histamplitude} shows that only $\sim10\%$ of the rotations identified in seasons~1--3 exceed the total range of amplitudes found in season~4.
The lack of such large amplitude rotations in season~4 may be due to small number statistics as only 9~rotations were identified.
The \ac{ad} indicates no significant difference between the two distributions of rotation amplitudes.\footnote{Amplitudes and durations are lower limits, when the rotations start or end at the start or end of an observing period. The results do not depend on whether or not we include the limits.}
A comparison of the distributions of durations, however, reveals a significant difference (\ac{ad} p-value $<0.001$).
\cref{fig:histduration} shows that $\gtrsim70\%$ of the rotations identified in seasons~1--3 have longer durations than the longest rotation in season~4.
Thus in season~4 we have identified none of the longer duration rotations that make up the majority of rotations in seasons~1--3.
We have also carried out this analysis separately for season~1, season~2, and season~3, {\it vs.\/} season~4, with the same result. 
In season~4 the cadence was $\sim7\times$ faster and the observing period was $\sim3\times$ shorter than in seasons~1--3.
The combination of both of these changes have likely led to the difference in long-duration rotations is season~4.

\subsubsection{Effects of shorter observing periods}
\label{sec:obsper}

\begin{table}
    \caption{
        Testing the effect of observing period length on the rotation identification.
        (1) Number of rotations. (2) Fraction of truncated rotations. (3) Mean of the ratio between rotation duration and corresponding observing period duration. (4) Occurrence rate of rotations per 100~days.}
    \label{tab:rotdur_vs_obsper}
    \centering
    \begin{tabular}{l c c c c}
        \toprule
        & Number    & Truncated    & $\overline{\rm rotation\,/}$ & Occurrence   \\
        & of        & rotations    & observing                    & per          \\
        & rotations & fraction$^*$ & period$^*$                   & 100 days$^*$ \\
        \midrule
        s1-3   &     26 & $  0.12\pm0.05$ & $  0.18\pm0.02$ &   0.25\\
        s4     &      9 & $  0.44\pm0.13$ & $  0.14\pm0.03$ &   0.73\\
        \bottomrule
        \multicolumn{5}{p{.9\columnwidth}}{$^*$Uncertainties are estimated with a bootstrap method; in 1000~iterations we select a random fraction of rotation events and repeat the analysis; for each measured property the uncertainty is given by the standard deviation of all bootstrap iterations.}
    \end{tabular}
\end{table}{}

Assuming the same underlying population of rotation events in seasons~1--3 and season~4, we expect three effects to be evident in season 4:

\begin{enumerate}
    \item Because the observing periods were shorter, we would expect more truncated rotations, i.e. rotations that start or end at the start or end of the observing periods. This is indeed what we find (\cref{tab:rotdur_vs_obsper}, col.~2).
    \item When rotations are not truncated the ratio between the rotation duration and the total observing period may be higher for season~4 than for seasons~1--3.
    We do not observe a significant difference (col.~3+4). For this analysis we excluded the truncated rotations.
    \item The intrinsic occurrence rate of rotations should not be affected by different observing period durations. However, shorter observing periods increase the chance of rotations falling on the edge of the period and the requirement of at least 4~data points for a detected rotation could decrease the number of identified rotations; but at the same time we have a faster observing cadence, which would counteract this effect.
    We observe that rotations occur about three times more frequently during season~4 than during seasons~1--3 (col.~5).
\end{enumerate}



\subsubsection{Effects of the observing cadence}
\label{sec:samplingrate}


Our definition of a rotation (c.f.~\cref{sec:samplingdependence}) identifies periods of data on different time separations that are similar in the sense that the \ac{evpa} changes are strong enough to produce a rotation larger than $90^\circ$ and smooth enough to be consistent with our requirement of smoothness.
As we have shown in \cref{fig:evpavartimescale}, the \ac{evpa} changes are generally smaller on shorter separations, such as the ones sampled during season~4, than on longer separations, such as the ones sampled during seasons~1--3.
As a consequence, during season~4 we may be picking out periods that are strongly variable and show faster rotation rates than seasons~1--3.
Furthermore, a faster cadence reveals shorter-timescale-variability.
The \ac{evpa} data do not show completely smooth trends, but vary on all separations.
A slower cadence may smooth out the shorter-timescale-\ac{evpa} changes to such an extent that smoother rotations are identified in more sparsely sampled data, which would not pass our smoothness criterion (c.f. \cref{sec:samplingdependence}) at a faster cadence.
As a consequence we would not identify rotations in season~4 having durations as long as those observed in seasons~1--3.
In fact, with the criterion of smoothness, we expect that some or all of the rotations identified in seasons~1--3 that have significantly longer durations than the rotations of season~4 would not have been identified as rotations if we had observed season~1--3 at faster cadence.


In summary we find that the identification of \ac{evpa} rotation candidates is strongly affected by cadence.
Therefore, results obtained from samples observed with substantially different cadences are not directly comparable, but must be analyzed carefully for the effects described above.
With a cadence substantially better than that of RoboPol seasons~1--3, our definition of smooth rotations may well need to be revised, since it appears that our requirement for smoothness is too restrictive and is therefore missing long-duration rotations.
More and faster cadence data are needed to make an informed decision whether \ac{evpa} rotations need to be defined and identified differently and, if that is the case, in particular what the revised smoothness criterion should be.

\section{Rotations}
\label{app:rotations}

\begin {figure*}
    \begin{minipage}{\textwidth}
    \centering
    \includegraphics[width=\textwidth]{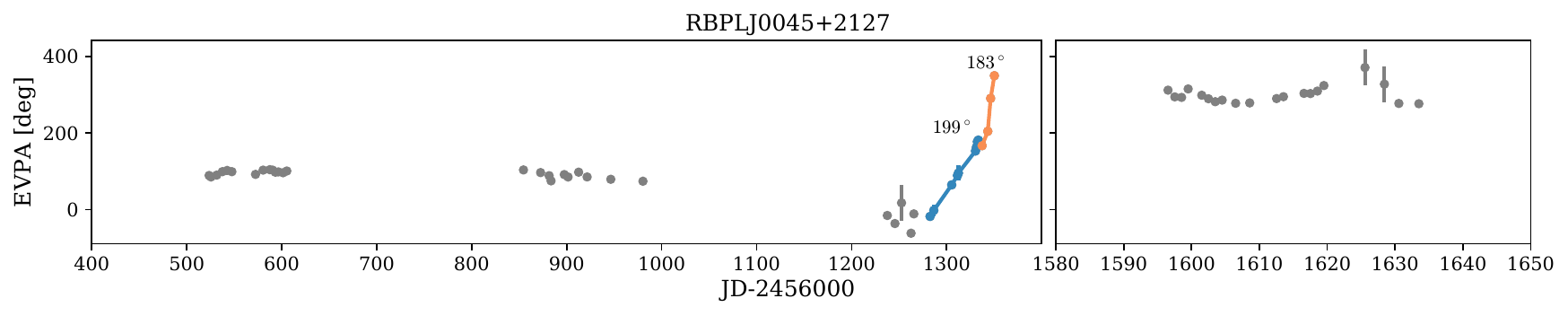}
    \end{minipage}
    \newline
    \begin{minipage}{\textwidth}
    \centering
    \includegraphics[width=\textwidth]{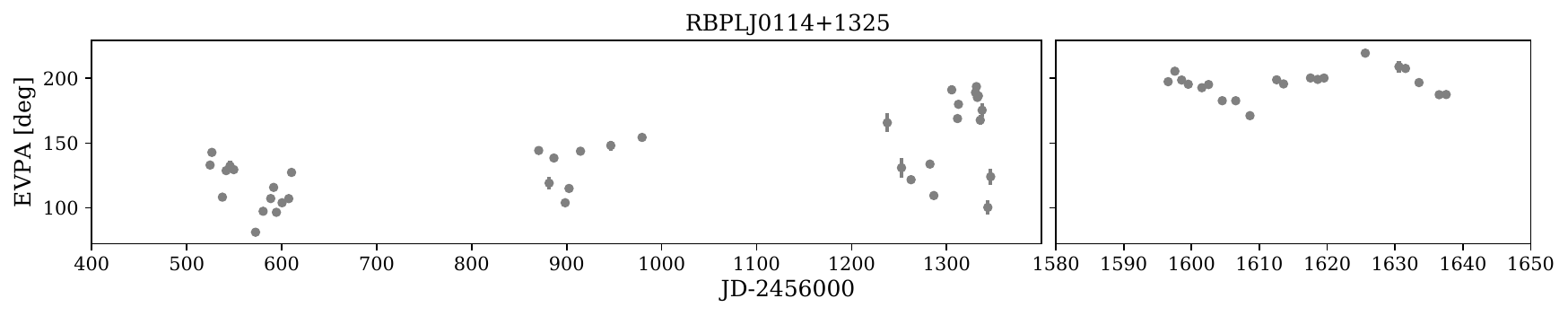}
    \end{minipage}
    \newline
    \begin{minipage}{\textwidth}
    \centering
    \includegraphics[width=\textwidth]{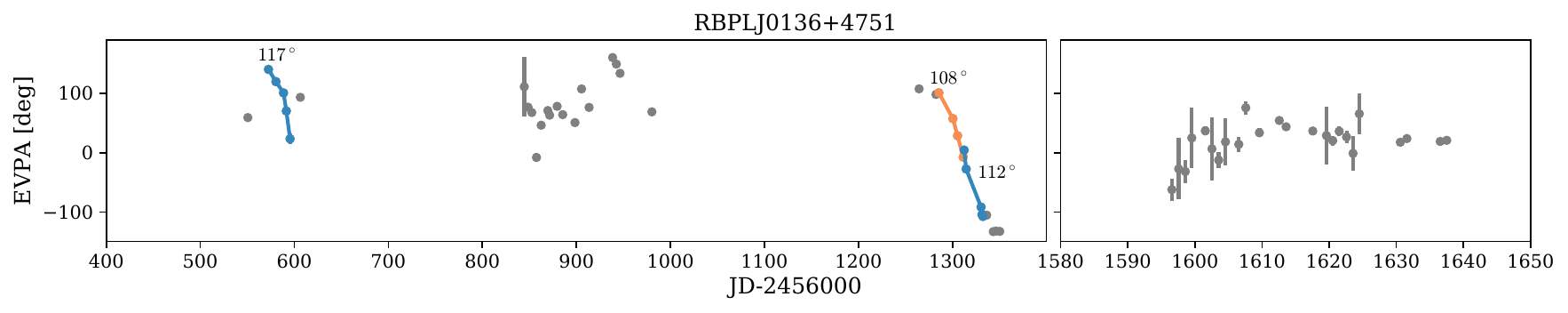}
    \end{minipage}
    \newline
    \begin{minipage}{\textwidth}
    \centering
    \includegraphics[width=\textwidth]{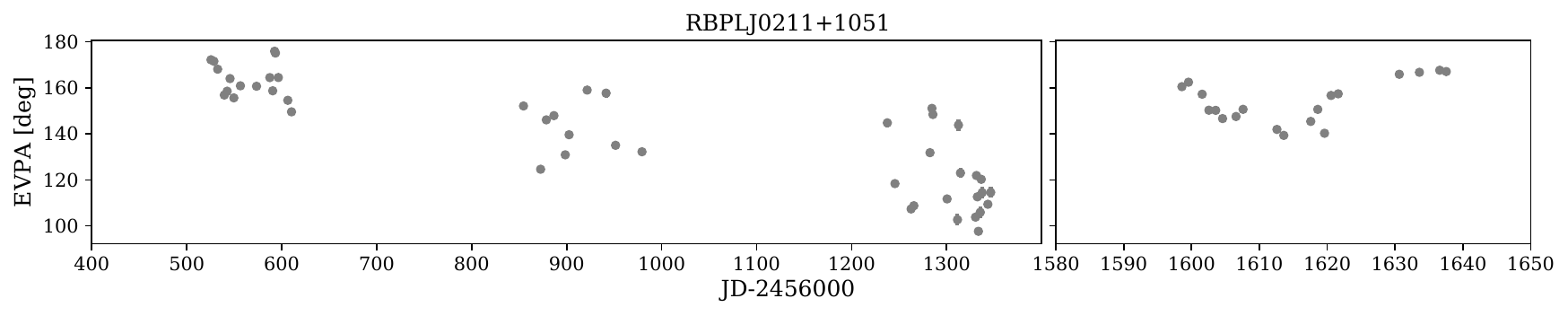}
    \end{minipage}
    \newline
    \begin{minipage}{\textwidth}
    \centering
    \includegraphics[width=\textwidth]{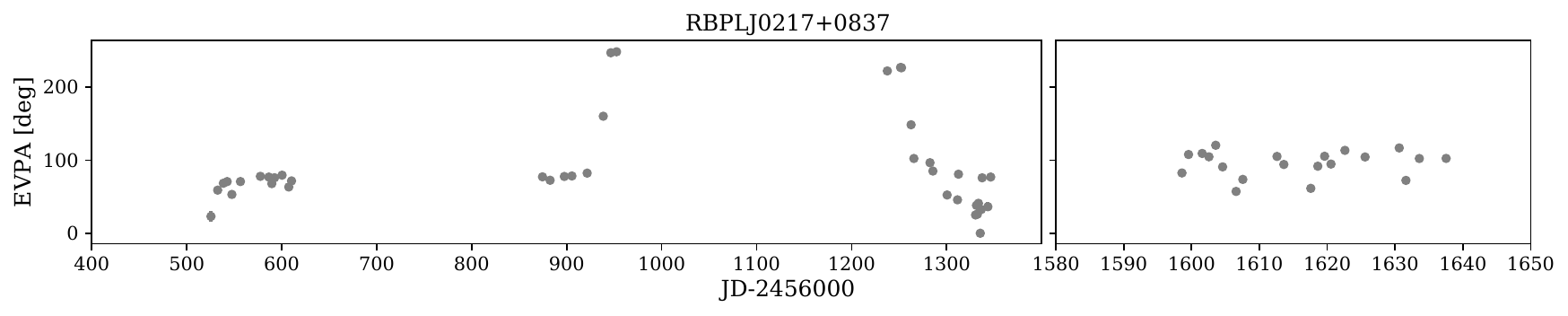}
    \end{minipage}
    \newline
    \begin{minipage}{\textwidth}
    \centering
    \includegraphics[width=\textwidth]{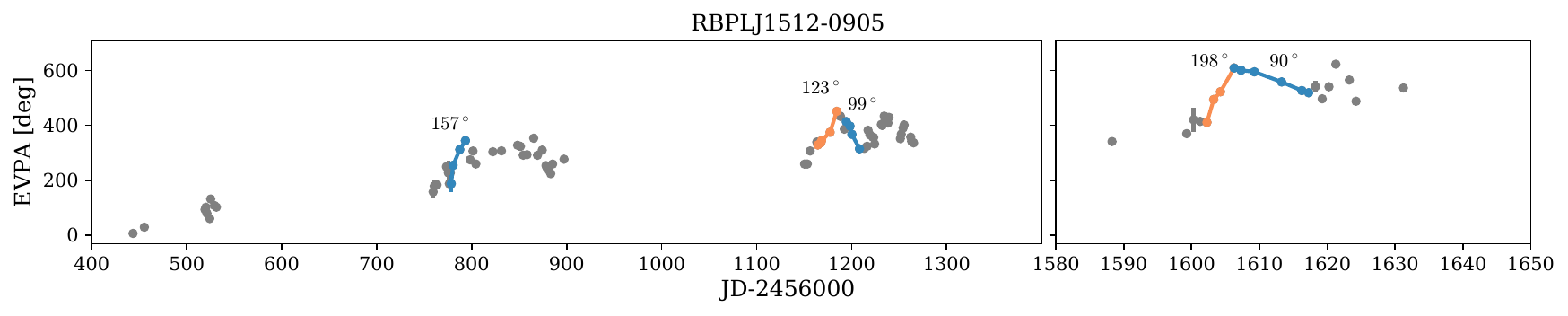}
    \end{minipage}
    \caption{Evolution of the adjusted \ac{evpa} over four seasons of observations of the RoboPol season~4 sample. The left panel shows seasons 1-3. The right panel shows season~4. Note that while the vertical scaling is the same in both panels, the horizontal axis scaling differs considerably between left and right panel. Coloured dots and lines highlight identified rotation periods. The colour alternates between blue and orange for a clearer visualization of different rotation periods.}
    \label{fig:rotations}
\end{figure*}

\Cref{fig:rotations} shows the evolution of the adjusted \ac{evpa} over four seasons of observations of the RoboPol season~4 sample. 
Coloured lines link data points that have been identified as rotations according to the criteria described in \cref{sec:samplingdependence}.
The amplitude of the identified rotations is written next to the rotations.
We note that some periods in the data may be identified as rotations by eye, but are not marked as such.
These periods are not consistent with the criteria that we described \cref{sec:samplingdependence}.
Typically, either the \ac{evpa} progression is not smooth enough or too few data points may have sampled the progression to be considered a rotation according to our strict criteria.


\renewcommand{\thefigure}{\thesection\arabic{figure} (continued)}
\addtocounter{figure}{-1}

\begin{figure*}
    \begin{minipage}{\textwidth}
    \centering
    \includegraphics[width=\textwidth]{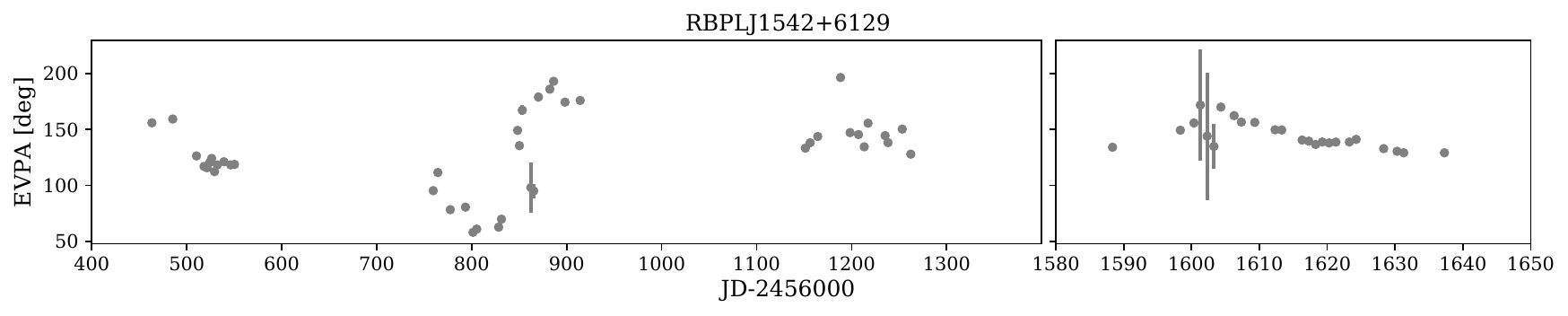}
    \end{minipage}
    \newline
    \begin{minipage}{\textwidth}
    \centering
    \includegraphics[width=\textwidth]{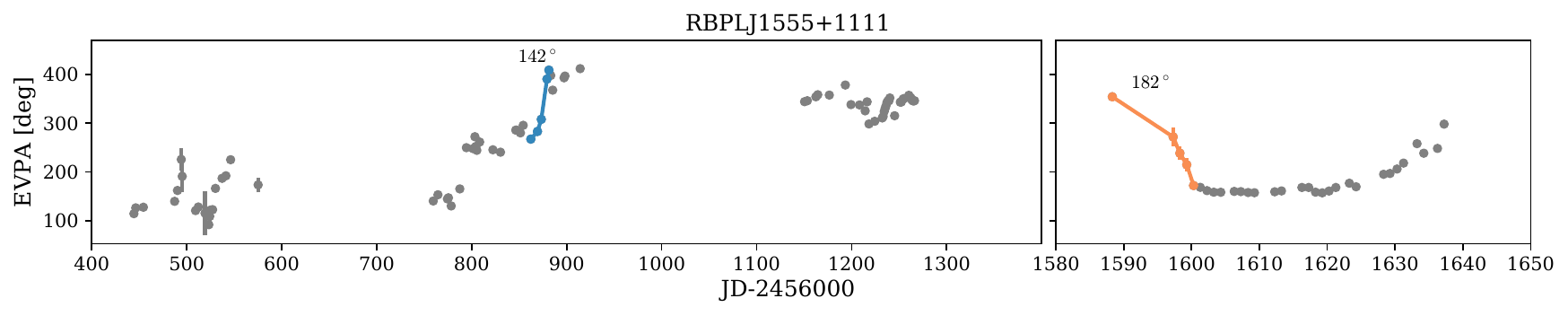}
    \end{minipage}
    \newline
    \begin{minipage}{\textwidth}
    \centering
    \includegraphics[width=\textwidth]{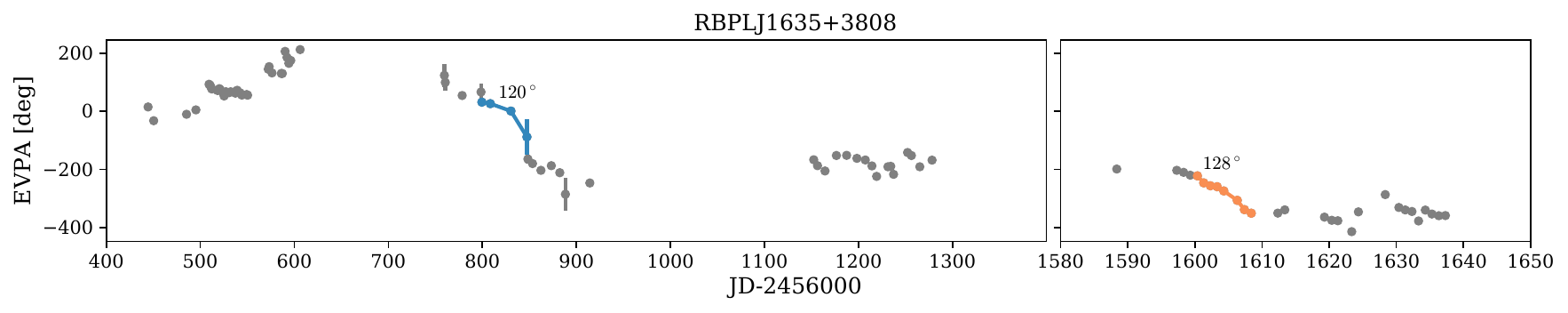}
    \end{minipage}
    \newline
    \begin{minipage}{\textwidth}
    \centering
    \includegraphics[width=\textwidth]{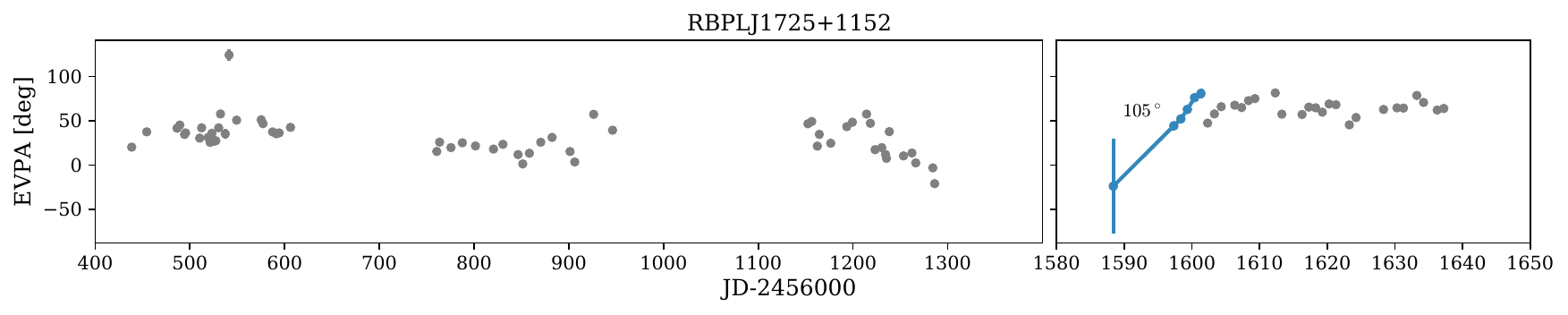}
    \end{minipage}
    \newline
    \begin{minipage}{\textwidth}
    \centering
    \includegraphics[width=\textwidth]{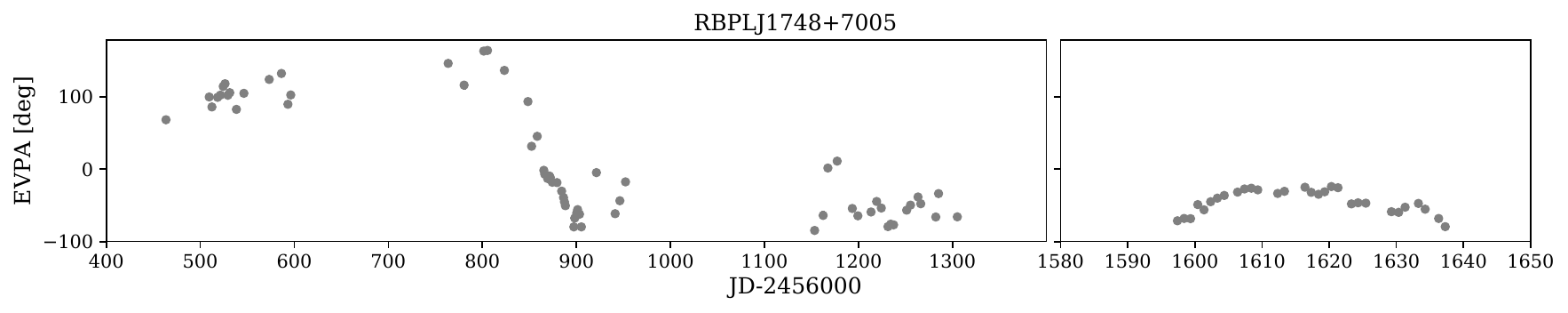}
    \end{minipage}
    \newline
    \begin{minipage}{\textwidth}
    \centering
    \includegraphics[width=\textwidth]{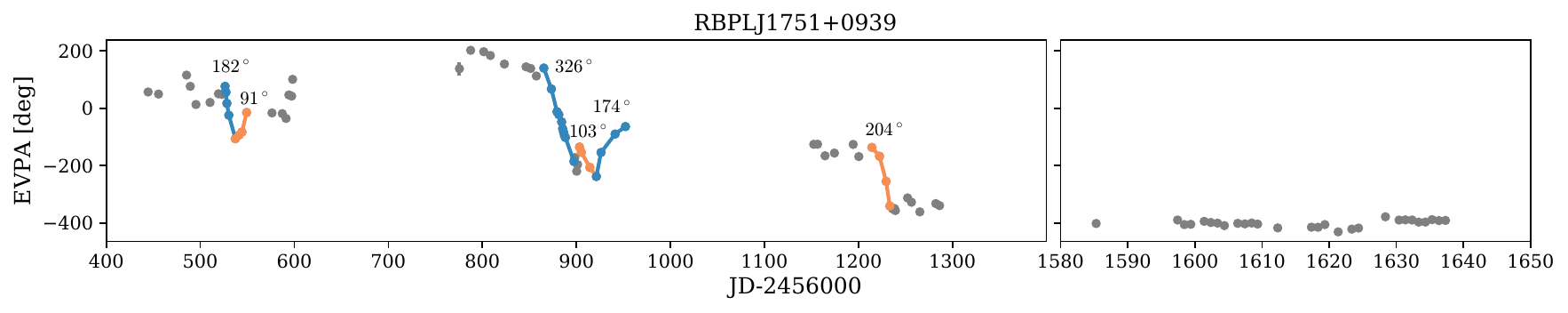}
    \end{minipage}
    \caption{}
\end{figure*}

\renewcommand{\thefigure}{\thesection\arabic{figure} (continued)}
\addtocounter{figure}{-1}

\begin{figure*}
    \begin{minipage}{\textwidth}
    \centering
    \includegraphics[width=\textwidth]{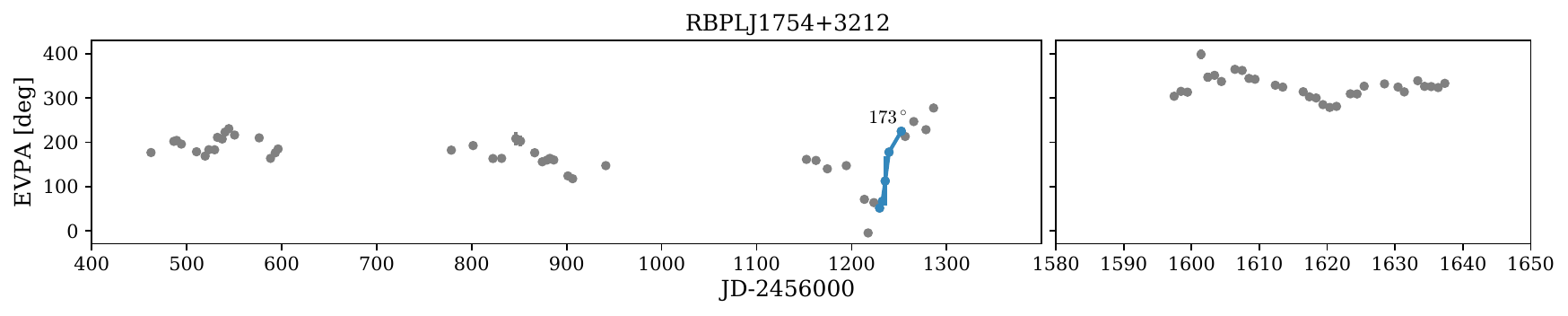}
    \end{minipage}
    \newline
    \begin{minipage}{\textwidth}
    \centering
    \includegraphics[width=\textwidth]{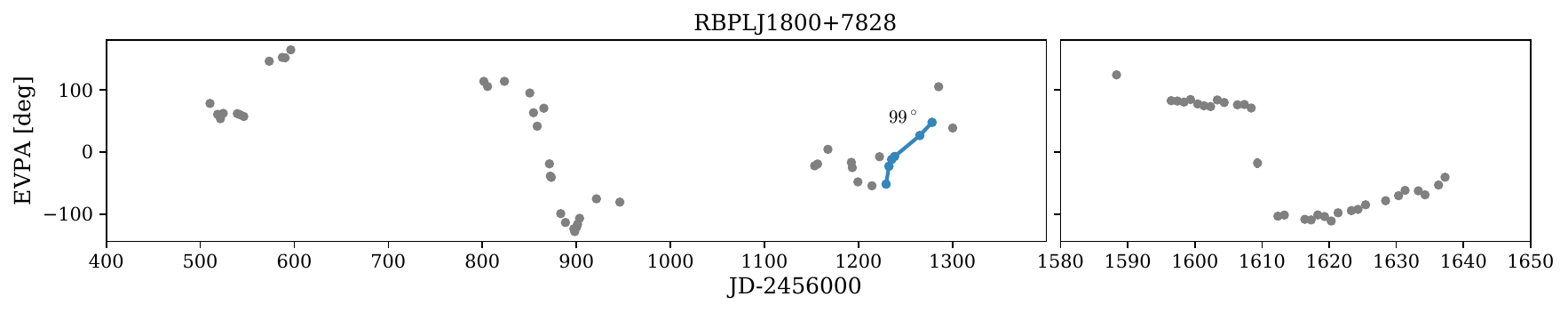}
    \end{minipage}
    \newline
    \begin{minipage}{\textwidth}
    \centering
    \includegraphics[width=\textwidth]{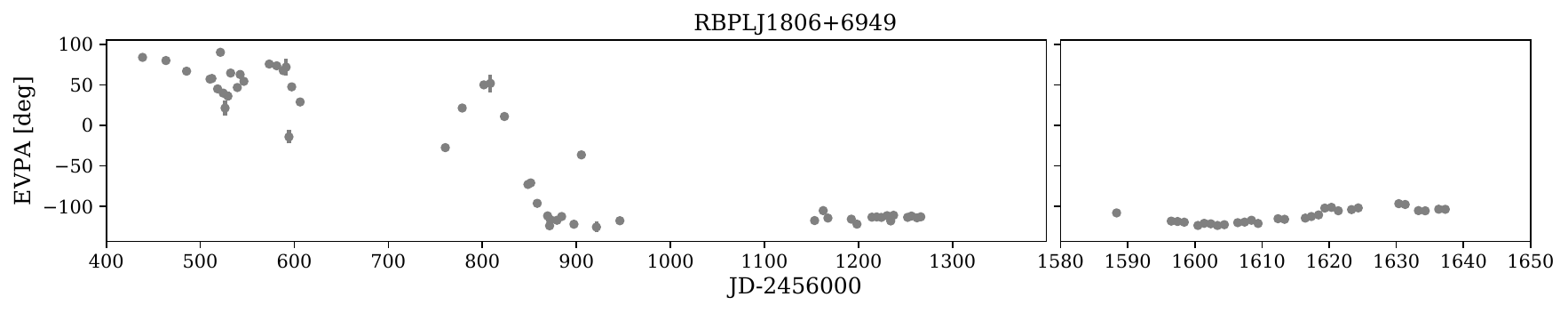}
    \end{minipage}
    \newline
    \begin{minipage}{\textwidth}
    \centering
    \includegraphics[width=\textwidth]{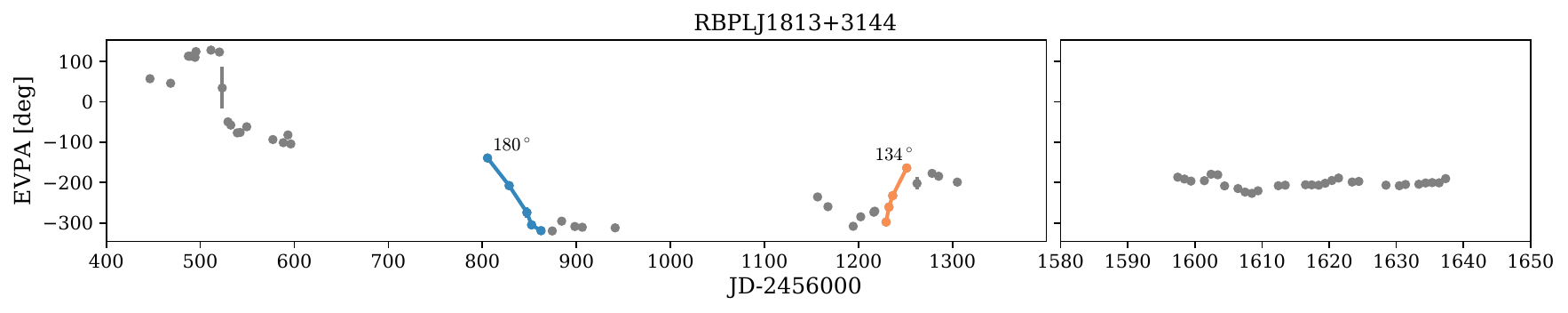}
    \end{minipage}
    \newline
    \begin{minipage}{\textwidth}
    \centering
    \includegraphics[width=\textwidth]{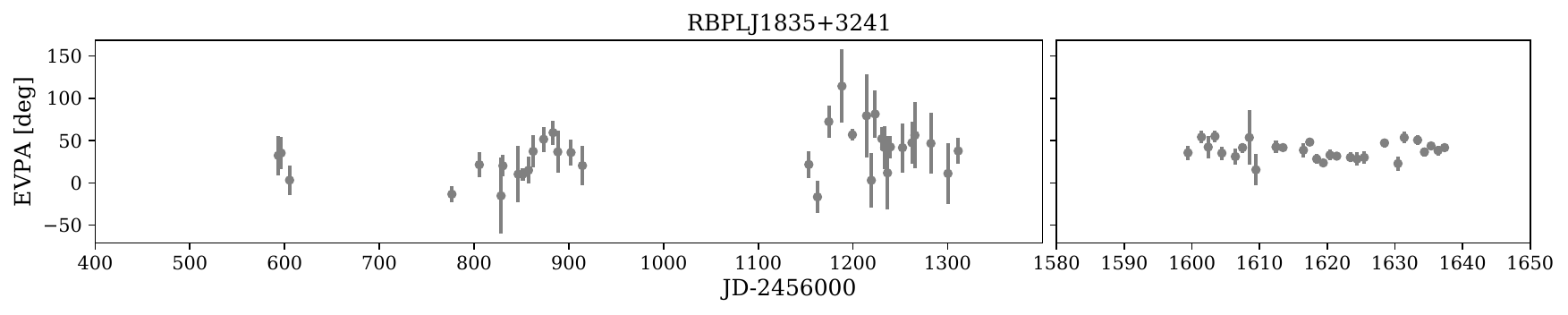}
    \end{minipage}
    \newline
    \begin{minipage}{\textwidth}
    \centering
    \includegraphics[width=\textwidth]{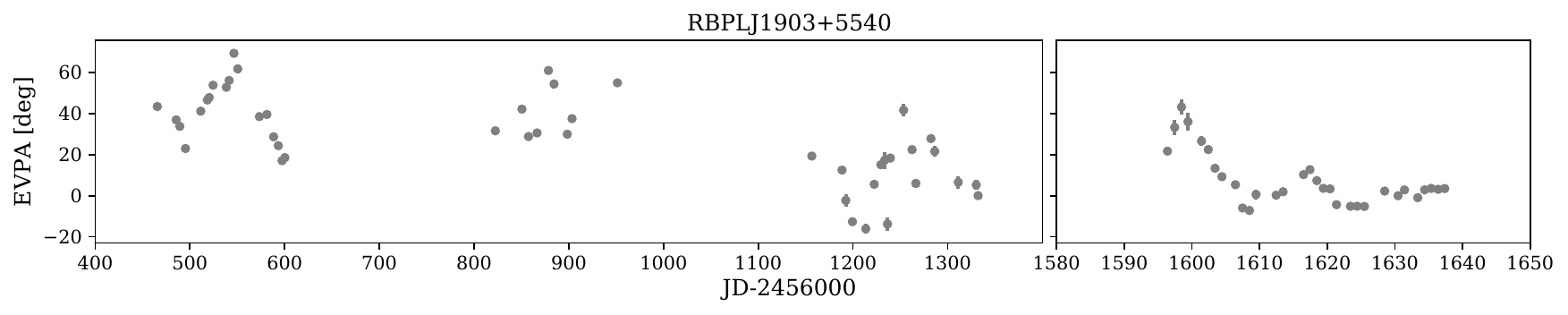}
    \end{minipage}
    \caption{}
\end{figure*}

\renewcommand{\thefigure}{\thesection\arabic{figure} (continued)}
\addtocounter{figure}{-1}

\begin{figure*}
    \begin{minipage}{\textwidth}
    \centering
    \includegraphics[width=\textwidth]{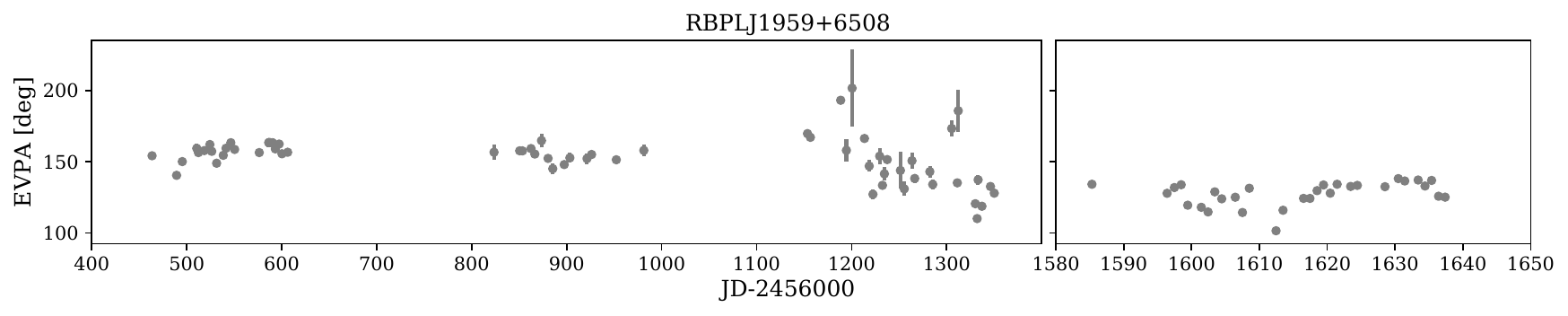}
    \end{minipage}
    \newline
    \begin{minipage}{\textwidth}
    \centering
    \includegraphics[width=\textwidth]{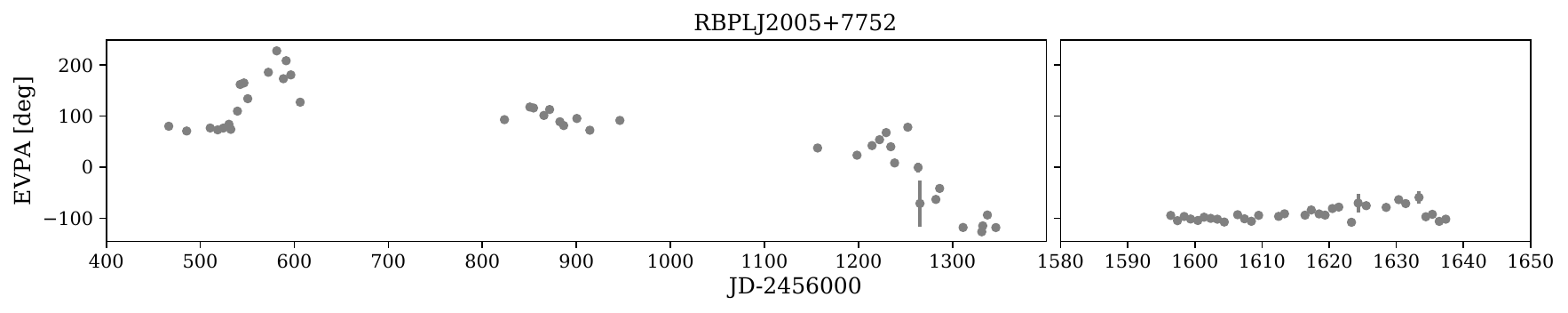}
    \end{minipage}
    \newline
    \begin{minipage}{\textwidth}
    \centering
    \includegraphics[width=\textwidth]{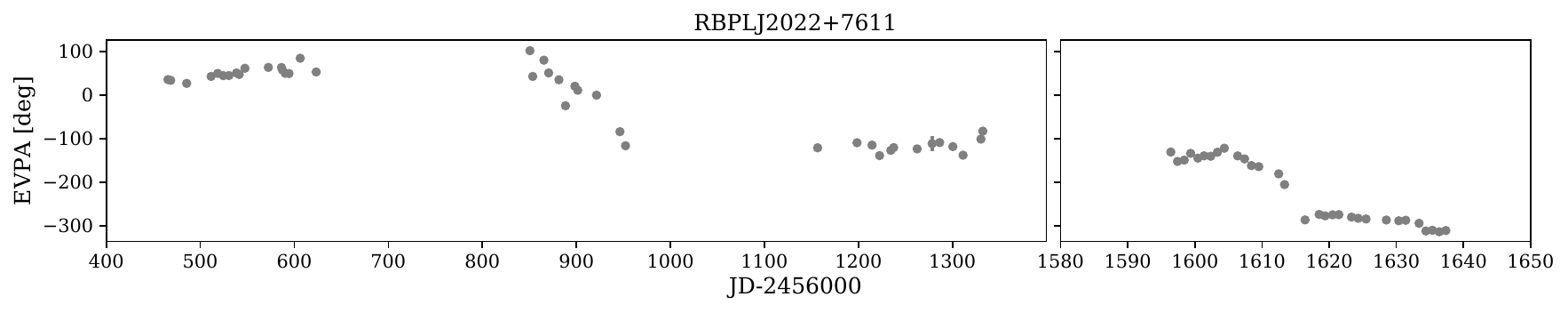}
    \end{minipage}
    \newline
    \begin{minipage}{\textwidth}
    \centering
    \includegraphics[width=\textwidth]{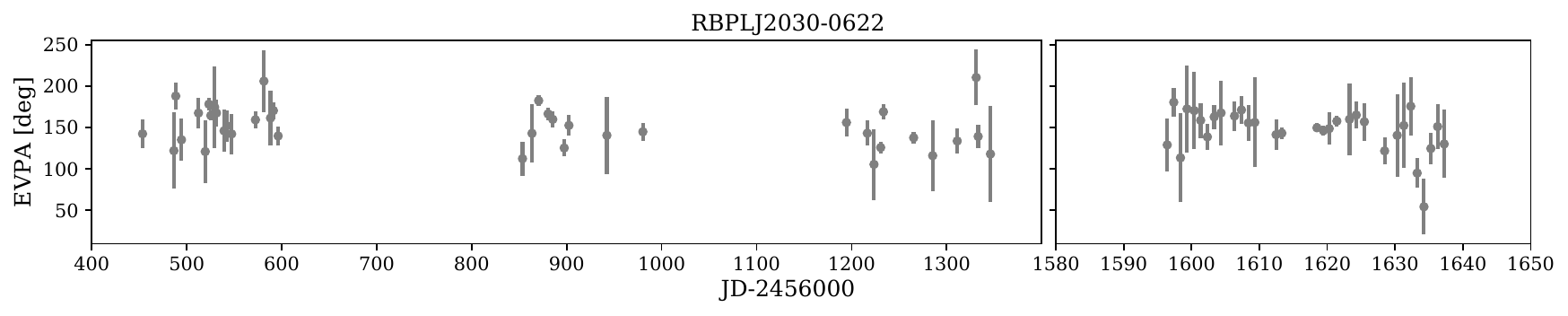}
    \end{minipage}
    \newline
    \begin{minipage}{\textwidth}
    \centering
    \includegraphics[width=\textwidth]{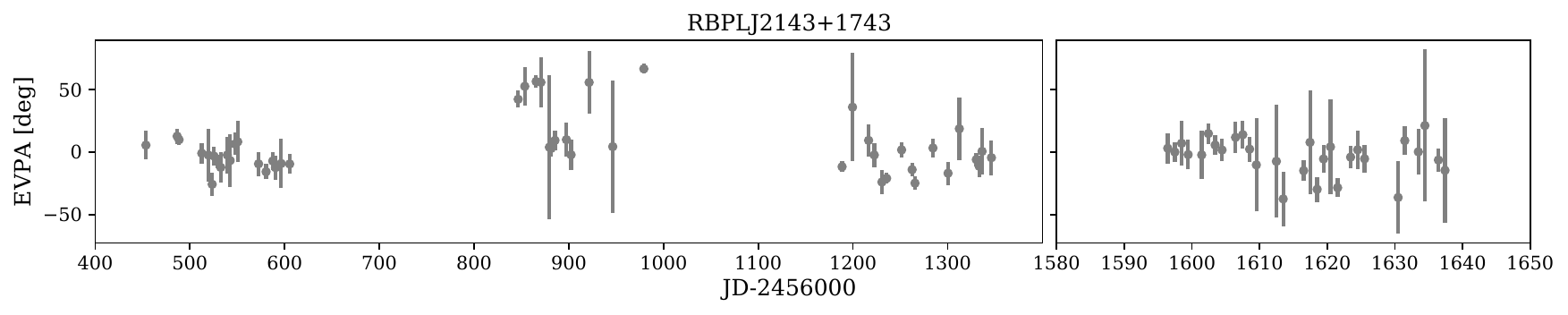}
    \end{minipage}
    \newline
    \begin{minipage}{\textwidth}
    \centering
    \includegraphics[width=\textwidth]{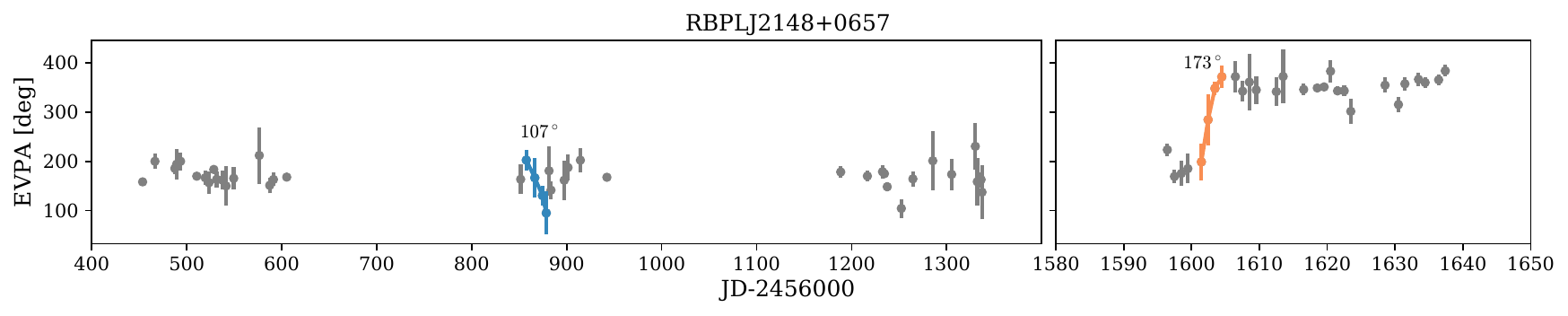}
    \end{minipage}
    \caption{}
\end{figure*}

\renewcommand{\thefigure}{\thesection\arabic{figure} (continued)}
\addtocounter{figure}{-1}

\begin{figure*}
    \begin{minipage}{\textwidth}
    \centering
    \includegraphics[width=\textwidth]{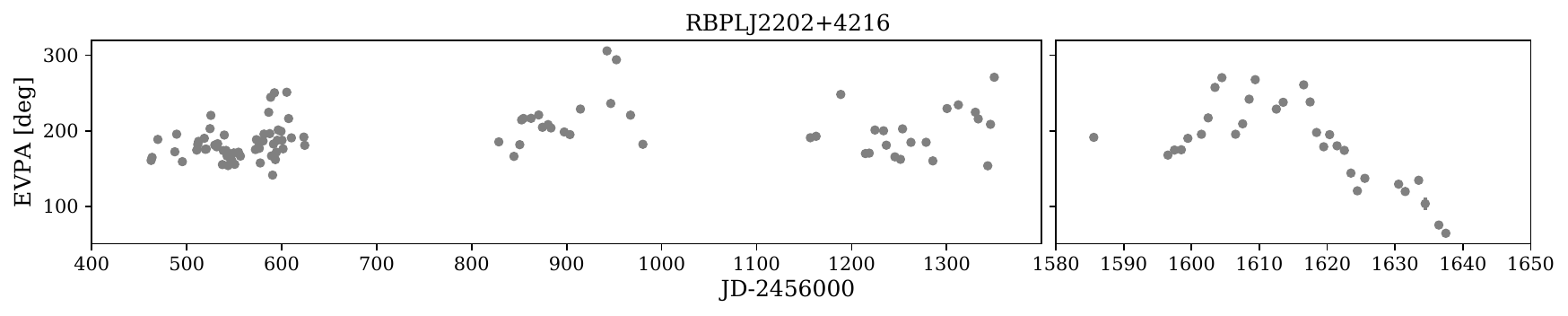}
    \end{minipage}
    \newline
    \begin{minipage}{\textwidth}
    \centering
    \includegraphics[width=\textwidth]{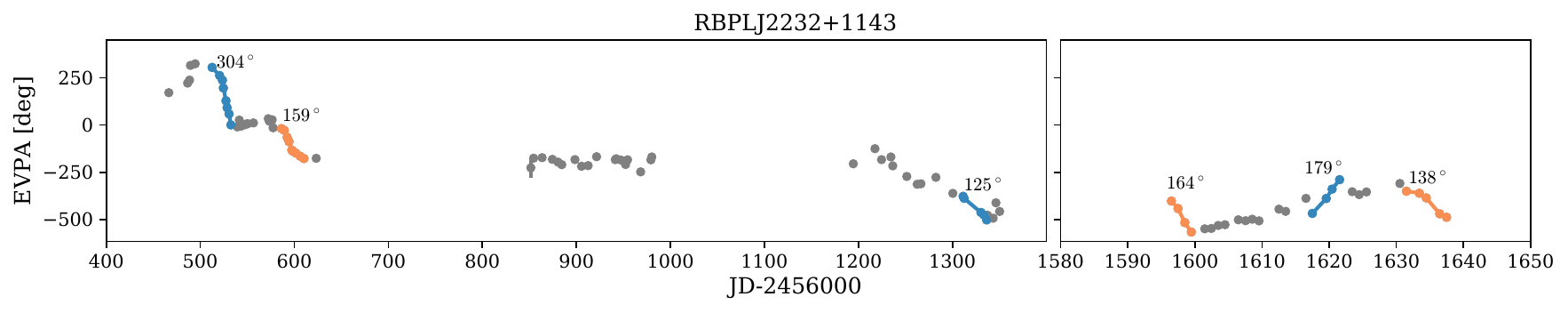}
    \end{minipage}
    \newline
    \begin{minipage}{\textwidth}
    \centering
    \includegraphics[width=\textwidth]{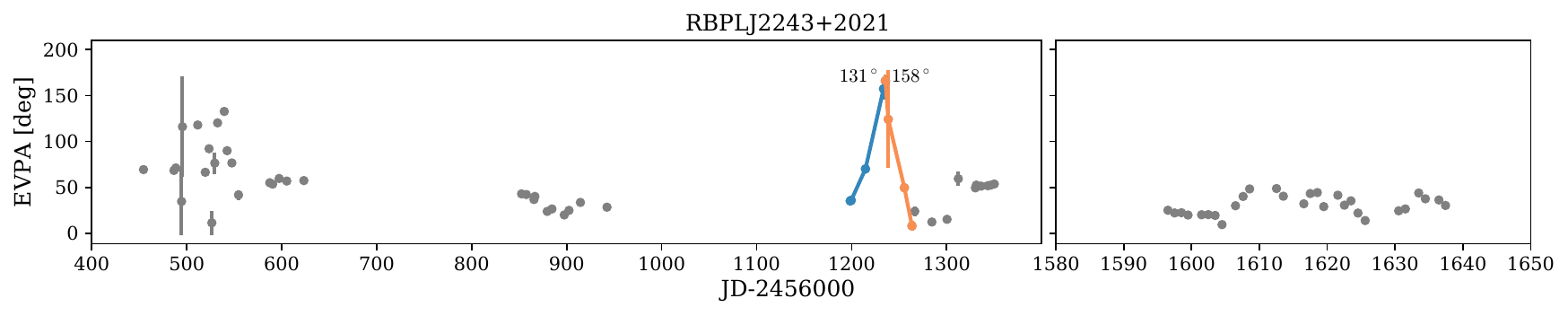}
    \end{minipage}
    \newline
    \begin{minipage}{\textwidth}
    \centering
    \includegraphics[width=\textwidth]{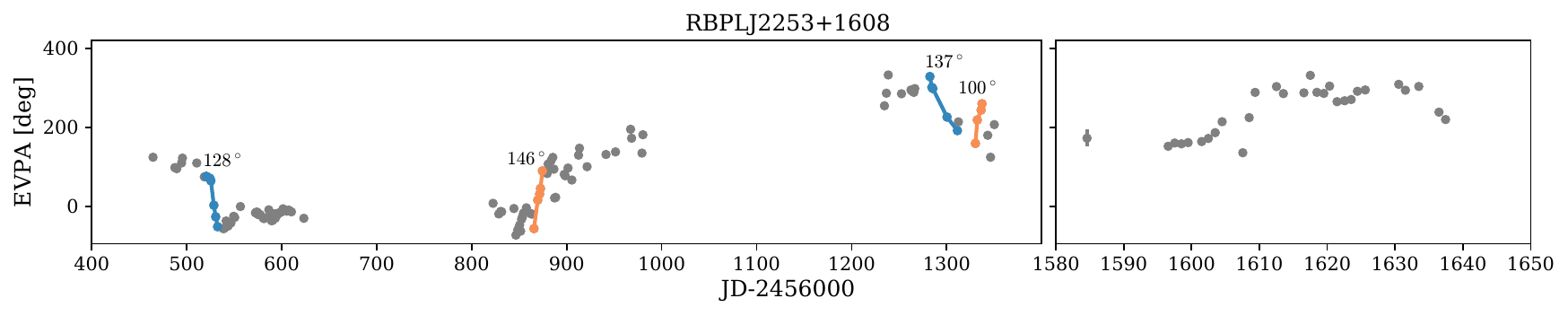}
    \end{minipage}
    \caption{}
\end{figure*}


\bsp	
\label{lastpage}
\end{document}